\documentclass[structabstract]{aa}

\usepackage[utf8]{inputenc}
\usepackage{graphicx}
\usepackage{natbib}
\usepackage[dvipsnames]{xcolor}
\usepackage{hyperref}
\usepackage{txfonts}
\usepackage{longtable}

\newcommand{\Xitau}{{\tt Xitau}}

\title{Towards a consistent model of the hot quadruple system HD\,93206 = QZ~Carin\ae: II. N-body model%
\thanks{Dedicated to the memory of Dr. Pavel Mayer.}
\thanks{Based on observations made with ESO Telescopes under programs 098.D-0706(B), 099.D-0777(B).}
}

\titlerunning{Towards a consistent model of the hot quadruple system HD\,93206 = QZ~Carin\ae: II. N-body model}

\author{
  M.~Bro\v{z}\inst{1}\and
  P.~Harmanec\inst{1}\and
  P.~Zasche\inst{1}\and
  R.~Catalan-Hurtado\inst{2,14}\and
  B.N.~Barlow\inst{2}\and
  W.~Frondorf\inst{2}\and
  M.~Wolf\inst{1}\and
  H.~Drechsel\inst{3}\and
  R.~Chini\inst{4,5}\and
  A.~Nasseri\inst{4}\and
  J.~Labadie-Bartz\inst{6}\and
  G.W.~Christie\inst{7}\and
  W.S.G.~Walker\inst{8}\and
  M.~Blackford\inst{9}\and
  D.~Blane\inst{10}\and
  A.A.~Henden\inst{11}\and
  T.~Bohlsen\inst{12}\and
  H.~Bo\v{z}i\'c\inst{13}\and
  J.~Jon\'ak\inst{1}
}

\institute{
  Charles University, Faculty of Mathematics and Physics, Institute of Astronomy, V~Hole{\v s}ovi{\v c}k{\'a}ch 2, CZ-18000 Prague, Czech Republic \and
  Department of Physics, High Point University, One University Way, High Point, NC 27268, USA \and
  Dr.~Karl~Remeis-Observatory \& ECAP, Astronomical Institute, Friedrich-Alexander-University Erlangen-Nuremberg, Sternwartstr.~7, 96049 Bamberg, Germany \and
  Astronomisches Institut, Ruhr-Universit\"at Bochum, Universit\"atsstr.~150, 44801 Bochum, Germany \and
  Instituto de Astronom\'\i a, Universidad Cat\'olica del Norte, Avenida Angamos 0610, Antofagasta, Chile \and
  Instituto de Astronomia, Geof\'\i sica e Ciencias Atmosf\'ericas, Universidade de S\~ao Paulo, Rua do Mat\~ao 1226, Cidade Universit\'aria, 05508-900 S\~ao Paulo, SP, Brazil \and
  Auckland Observatory, PO Box 24180, Royal Oak, Auckland, New Zealand \and
  Variable Stars South, P O Box 173, Awanui, New Zealand, 0451 \and
  Variable Stars South, Congarinni Observatory, Congarinni, NSW, Australia~2447 \and
  Variable Stars South, and Astronomical Society of Southern Africa, Henley Observatory, Henley on Klip, Gautenburg, South Africa \and
  AAVSO, 106 Hawking Pond Road, Center Harbor, NH03226, USA \and
  Mirranook Observatory, Boorolong Rd Armidale, NSW, 2350, Australia \and
  Hvar Observatory, Faculty of Geodesy, Zagreb University, Ka\v{c}i\'ceva~26, 10000 Zagreb, Croatia \and
  Eukaryotic Pathogens Innovation Center, Clemson University, Clemson, South Carolina 29634, USA
}

\date{Received x-x-2022 / Accepted x-x-2022}

\abstract
   {}
   {
HD\,93206 is early-type massive stellar system,
composed of components resolved by direct imaging (Ab, Ad, B, C, D)
as well as a compact sub-system (Aa1, Aa2, Ac1, Ac2).
Its geometry was already determined on the basis of extensive
photometric, spectroscopic and interferometric observations.
However, the fundamental absolute parameters are still not known
precisely enough.
   } 
   {
We use an advanced N-body model to account for
all mutual gravitational perturbations among the four close components, and
all observational data types, including:
astrometry,
radial velocities,
eclipse timing variations,
squared visibilities,
closure phases,
triple products,
normalized spectra, and
spectral-energy distribution (SED).
The respective model has 38 free parameters, namely
three sets of orbital elements, component masses, and their basic radiative properties ($T$, $\log g$, $v_{\rm rot}$).
   }
   {
We revised the fundamental parameters of QZ~Car as follows.
For a model with the nominal extinction coefficient $R_V \equiv A_V/E(B-V) = 3.1$,
the best-fit masses are
$m_1 = 26.1\,M_{\rm S}$,
$m_2 = 32.3\,M_{\rm S}$,
$m_3 = 70.3\,M_{\rm S}$,
$m_4 = 8.8\,M_{\rm S}$,
with uncertainties of the order of $2\,M_{\rm S}$,
and the system distance
$d = (2800\pm 100)\,{\rm pc}$.
In an alternative model, where we increased the weights
of RV and TTV observations and relaxed the SED constraints,
because extinction can be anomalous with $R_V \sim 3.4$,
the distance is smaller, $d = (2450\pm 100)\,{\rm pc}$.
This would correspond to that of Collinder~228 cluster.
Independently, this is confirmed by dereddening of the SED,
which is only then consistent with the early-type classification
(O9.7Ib for Aa1, O8III for Ac1).
Future modelling should also account for an accretion disk
around Ac2 component.
   }
   {}
 

\begin{document}

\maketitle

%%%%%%%%%%%%%%%%%%%%%%%%%%%%%%%%%%%%%%%%%%%%%%%%%%%%%%%%%%%%%%%%%%%%%%%%

\section{Introduction}

HD\,93206 is a complex system composed of nine components
(usually denoted as Aa1, Aa2, Ac1, Ac2, Ab, Ad, B, C, D;
\citealt{Mason_2001AJ....122.3466M}).
In the accompanying paper \cite{Mayer_2022arXiv220407045H}, we used several
observation-specific models to obtain fundamental parameters
of its sub-systems,
in particular the eclipsing binary Ac1+Ac2,
the spectroscopic binary Aa1+Aa2,
and their mutual long-period orbit (Aa1+Aa2)+(Ac1+Ac2).
Together, the quadruple sub-system is denoted as QZ~Car variable star.
All these models were kinematical,
i.e., based the two-body problem
with the light-time effect
and the third light,
to account at least for major perturbations.

In this work, we focus on the same quadruple sub-system QZ~Car,
but we use a dynamical N-body model called \Xitau%
\footnote{\url{http://sirrah.troja.mff.cuni.cz/~mira/xitau/}},
to account for additional perturbations (in particular, all mutual,
parametrized post-Newtonian).
Moreover, we account for all observational datasets at the same time, including
astrometry,
photometry,
spectroscopy,
interferometry,
etc.
This approach allows us to derive robust estimates of fundamental parameters.
Our model was already used for multiple stellar systems
(e.g, $\xi$~Tau, \citealt{Nemravova_2016A&A...594A..55N,Broz_2017ApJS..230...19B})
and asteroidal moons systems
with highly-irregular central bodies
(e.g., (216), \citealt{Marchis_2021A&A...653A..57M,Broz_2021A&A...653A..56B,Broz_2022A&A...657A..76B}).

%%%%%%%%%%%%%%%%%%%%%%%%%%%%%%%%%%%%%%%%%%%%%%%%%%%%%%%%%%%%%%%%%%%%%%%%

\section{Observational datasets}

While radial velocity measurements,
relative spectroscopy,
light curves and corresponding eclipse-timing variations (ETVs), and
astrometry
were already described in \cite{Mayer_2022arXiv220407045H},
here we focus on additional datasets,
namely the spectral-energy distribution and interferometry.

\subsection{Spectral-energy distribution (SED)}\label{sed}

We used data from Vizier,
which included the standard Johnson system photometry \citep{Ducati_2002yCat.2237....0D},
verified by La Silla and Sutherland photometry (see \citealt{Mayer_2022arXiv220407045H}),
Hipparcos \citep{Anderson_2012AstL...38..331A},
GAIA3 \citep{Gaia_2020yCat.1350....0G},
2MASS \citep{Cutri_2003yCat.2246....0C},
WISE \citep{Cutri_2012yCat.2311....0C},
MSX \citep{Egan_2003yCat.5114....0E}, and
Akari \citep{Ishihara_2010A&A...514A...1I}.
Altogether, the whole spectral range is $0.35$ to $22\,\mu{\rm m}$.

\begin{figure}
\centering
\includegraphics[width=8.5cm]{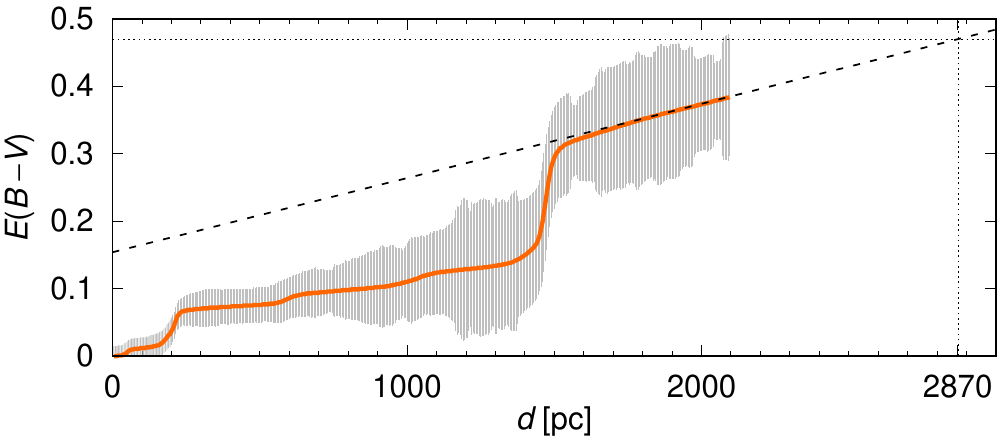}
\caption{
Reddening $E(B-V)$ vs. distance for the direction towards QZ~Car,
according to \cite{Lallement_2019A&A...625A.135L}.
The mean (\color{orange}orange\color{black}),
dispersion (\color{gray}gray\color{black}),
and extrapolation (dashed)
to the nominal distance 2870\,pc are shown.
}
\label{gaiadr2_2870pc}
\end{figure}

\begin{figure}
\centering
\includegraphics[width=8.5cm]{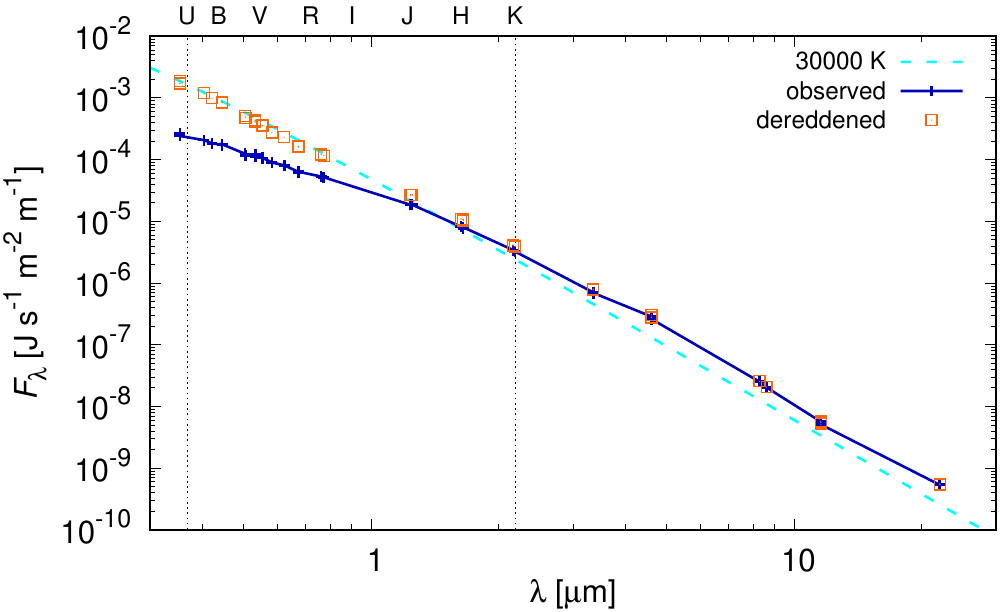}
\caption{
Observed spectral-energy distribution (SED; \color{blue}blue\color{black})
and dereddened SED (\color{orange}orange\color{black})
for the distance 2870\,pc.
The dereddened fluxes $F_\lambda$ approximately correspond to
black-body radiation with the temperature 30\,000\,K
(\color{cyan}cyan\color{black}).
In far infrared, where absorption is negligible,
a non-negligible (half order) excess is present.
}
\label{Sed_}
\end{figure}

It was necessary to carry out dereddening of these data.
Unfortunately, classic extinction maps of \cite{Green_2019ApJ...887...93G}
give unrealistic absorption
and a special treatment is needed for QZ~Car,
which is too close to the galactic equator.
To this point, we used extinction maps of \cite{Lallement_2019A&A...625A.135L}.
For the assumed distance 2870\,pc \citep{Shull_2019ApJ...882..180S},
the reddening extrapolated from 2100\,pc is
$E(B-V) = (0.470 \pm 0.060)\,{\rm mag}$,
and corresponding absorption
$A_V = 3.1\,E(B-V) = (1.46\pm 0.19)\,{\rm mag}$
(Fig.~\ref{gaiadr2_2870pc}).
Using the standard wavelength dependence of \cite{Schlafly_2011ApJ...737..103S},
we obtained the dereddened fluxes,
shown in Fig.~\ref{Sed_}.

Not surprisingly, the corrected fluxes exhibit a power-law slope,
which corresponds to the Planck function
($B_\lambda \propto \lambda^{-4}$ for $\lambda > 0.35\,\mu{\rm m}$),
for the effective temperature $T \simeq 30000\,{\rm K}$.
This is in agreement with the early-type spectral classification of QZ~Car.
Using a significantly smaller distance (e.g., 2000\,pc)
would result in a SED which does not correspond to early-type stars.
We shall see later in Section~\ref{sec:model}
that this distance and dereddening are indeed reliable.
On the other hand, in far infrared (${>}\,2.2\,\mu{\rm m}$),
where absorption is already negligible,
a non-negligible excess is present.
The monochromatic fluxes are a factor of 5 too large,
possibly indicating cold circumstellar matter or blending,
neither of which is accounted for in our modelling.
Consequently, we should preferably fit the NUV--NIR spectral range.

%%%%%%%%%%%%%%%%%%%%%%%%%%%%%%%%%%%%%%%%%%%%%%%%%%%%%%%%%%%%%%%%%%%%%%%%

\subsection{Interferometry}\label{interferometry}

Additionally to \cite{Sanchez_2017ApJ...845...57S} astrometric data,
for the epochs Jun 17th 2016, Jun 18th 2016,
we used interferometric observations with the VLTI/GRAVITY instrument,
found in the ESO archive for the epochs
Mar 14th 2017,
Apr 27th 2017.
This is an important constraint for the (Ac1+Ac2)+(Aa1+Aa2) orbit.
Apart from astrometry, we can directly include
squared visibilities $V^2$,
closure phases ${\rm arg} T_3$, and
triple product amplitudes $|T_3|$
in~\Xitau.

\begin{figure}
\centering
\includegraphics[width=9cm]{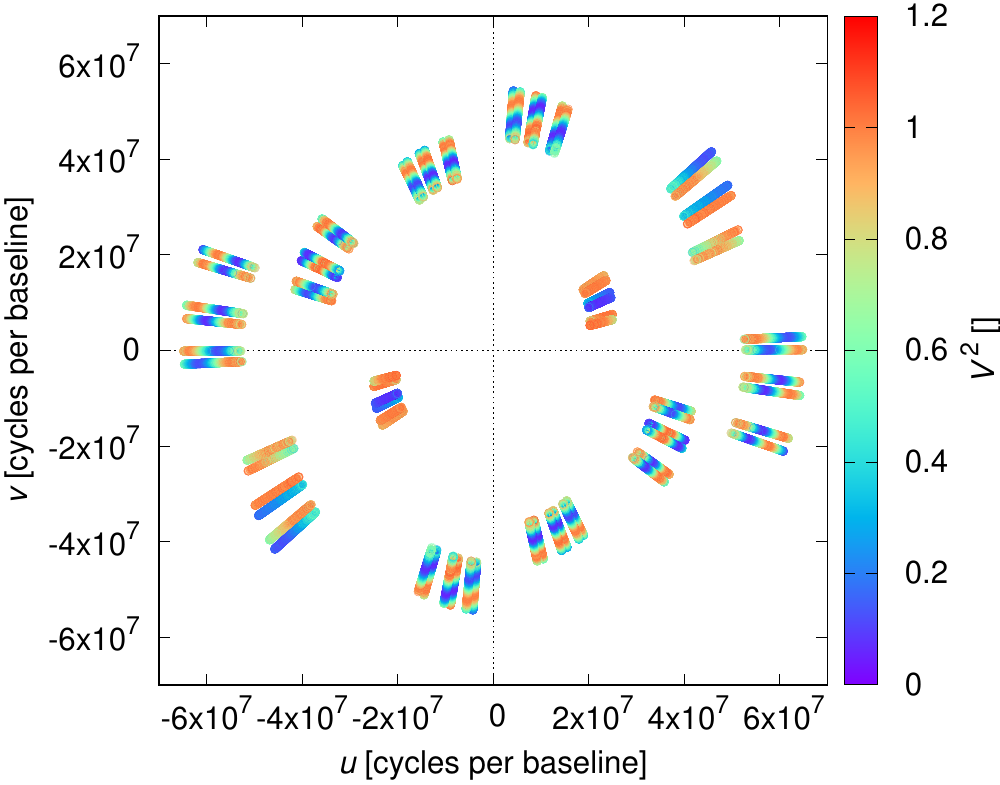}
\caption{
Coverage $(u, v) \equiv \vec B/\lambda$ (in cycles per baseline)
of new interferometric observations (see Tab.~\ref{tab:litpro}).
The corresponding squared visibility $V^2$ is plotted in colour.
The wide orbit ((Ac1+Ac2)+(Aa1+Aa2)),
having the angular separation more than $30\,{\rm mas}$,
is clearly resolved.
}
\label{uv}
\end{figure}

We used the Esoreflex VLTI/GRAVITY pipeline
\citep{Freudling_2013A&A...559A..96F,Abuter_2017A&A...602A..94G}
to reduce the data.
We performed the standard visibility calibration,
which is substantial (0.2) for baselines $\ge 4.5\times 10^7$ cycles per baseline.
We used the so-called `vfactor' for the correction of $V^2$.
We did not use data with $V^2 + \sigma_{V^2} > 1.0$.
The new $(u, v)$ coverage is shown in Fig.~\ref{uv}.
The wide orbit ((Ac1+Ac2)+(Aa1+Aa2)),
having the angular separation more than $30\,{\rm mas}$,
is clearly resolved.
On the other hand, angular diameters of individual components
($0.015$ to $0.08\,{\rm mas}$) cannot be resolved.
We shall see later in Section~\ref{sec:model},
there were some problems with $|T_3|$ calibration.
Consequently, we also tried to reduce the data without `vfactor'.

As a check, we performed astrometry with the Litpro software
\citep{TallonBosc_2008SPIE.7013E..1JT}.
A simplified model of a binary (Ac+Aa) was assumed.
We fitted $(u, v)$ coordinates of Aa and the flux ratio $F_{\rm Aa}/F_{\rm Ac}$;
other parameters were kept fixed.
Even though there are some systematics,
the fit converges without problems and astrometric uncertainties
are comparable to previous measurements ($\sigma_{u, v} \simeq 0.010\,{\rm mas}$;
Tab.~\ref{tab:litpro}).%
\footnote{In \Xitau, (Ac1+Ac2) eclipsing binary is located in the centre,
hence we use negative signs of $(u, v)$ for computations.}

\vskip\baselineskip
We used no light curves in our modelling (only indirectly as ETVs).
To our knowledge, there are no sufficiently high-signal-to-noise
differential interferometric measurements
(cf.~\cite{Sanchez_2017ApJ...845...57S}),
which would otherwise constrain velocity-dependent visibilities
across spectral lines.

\begin{table*}
\centering
\caption{Astrometric positions of component (Ac1+Ac2)
with respect to (Aa1+Aa2) from literature and new data.}
\begin{tabular}{rrrrrrrl}
\hline
time  & $u$ & $v$ & major & minor & P.A. & $F_{\rm Ac}/F_{\rm Aa}$ & Ref. \\ 
(UTC) & mas & mas & mas   & mas   & deg  & 1 \\
\hline
2012.1241           & 13.590 & 24.538 & 5.410 & 4.705 & 331.020 & $0.68\pm0.10$ & \cite{Sana_2014ApJS..215...15S} \\
2012.4409           & 12.367 & 22.597 & 0.540 & 0.652 & 331.310 & $0.37\pm0.04$ & \cite{Sana_2014ApJS..215...15S} \\
2016-06-18T01:35:21 & 15.940 & 25.609 & 0.045 & 0.054 & 0       & $0.74\pm0.14$ & \cite{Sanchez_2017ApJ...845...57S} \\
2017-03-14T03:34:17 & 16.370 & 25.672 & 0.003 & 0.003 & 0       & $0.490\pm0.001$ & ESO archive 098.D-0706(B) \\
2017-04-27T01:20:12 & 16.390 & 25.664 & 0.003 & 0.003 & 0       & $0.571\pm0.001$ & ESO archive 099.D-0777(B) \\
2017-04-27T02:08:58 & 16.402 & 25.678 & 0.003 & 0.003 & 0       & $0.573\pm0.001$ & ESO archive 099.D-0777(B) \\
\hline
\end{tabular}
\label{tab:litpro}
\end{table*}

%%%%%%%%%%%%%%%%%%%%%%%%%%%%%%%%%%%%%%%%%%%%%%%%%%%%%%%%%%%%%%%%%%%%%%%%

\section{Dynamical N-body model}\label{sec:model}

In \Xitau, the dynamical model contains the following accelerations terms
\citep{Broz_2017ApJS..230...19B,Broz_2021A&A...653A..56B,Broz_2022A&A...657A..76B}:
\begin{equation}
\vec f = -\sum_{j\ne i}^N {Gm_j\over r_{ij}^3}\vec r_{ij} + \vec f_{\rm ppn} + \vec f_{\rm oblat} + \vec f_{\rm multipole} + \vec f_{\rm tides}\,,
\end{equation}
where the terms are mutual gravitational interactions
and the parametrized post-Newtonian (PPN);
we do not activate oblateness, multipoles or tides for QZ~Car.
We superseded the PPN approximation by that of \cite{Standish_2006}, which is more accurate.
Hereinafter, the notation conforms to the actual implementation:
\begin{eqnarray}
\vec f_{\rm ppn} &=& \sum_{j\ne i}^N\,\bigl[-K_1\left(K_2+K_3+K_4+K_5+K_6+K_7+K_8\right)\vec r_{ij}+ \\
&&+\, K_1\left(K_9+K_{10}\right)\dot{\vec r}_{ij} + K_{11}\ddot{\vec r}_j\bigr]\,,
\end{eqnarray}
\begin{equation}
K_1 = {1\over c^2}{Gm_j\over r_{ij}^3}\,,
\end{equation}
\begin{equation}
K_2 = -2(\beta+\gamma)\sum_{k\ne i}{Gm_k\over r_{ik}}\,,
\end{equation}
\begin{equation}
K_3 = -(2\beta-1)\sum_{k\ne j}{Gm_k\over r_{jk}}\,,
\end{equation}
\begin{equation}
K_4 = \gamma v_i^2\,,
\end{equation}
\begin{equation}
K_5 = (1+\gamma) v_j^2\,,
\end{equation}
\begin{equation}
K_6 = -2(1+\gamma)\,\dot{\vec r}_i\!\cdot\dot{\vec r}_j\,,
\end{equation}
\begin{equation}
K_7 = -{3\over 2}{(\vec r_{ij}\cdot\dot{\vec r}_j)^2\over r_{ij}^2}\,,
\end{equation}
\begin{equation}
K_8 = {1\over 2} \vec r_{ji}\cdot\ddot{\vec r}_j\,,
\end{equation}
\begin{equation}
K_9 = (2+2\gamma)\,\vec r_{ij}\!\cdot\dot{\vec r}_i\,,
\end{equation}
\begin{equation}
K_{10} = -(1+2\gamma)\,\vec r_{ij}\!\cdot\dot{\vec r}_j\,,
\end{equation}
\begin{equation}
K_{11} = {3+4\gamma\over 2c^2}{Gm_j\over r_{ij}}\,,
\end{equation}
where
$\vec r_i$ denotes the position vector of body~$i$,
$\vec r_{ij} \equiv \vec r_i-\vec r_j$,
$v_i \equiv |\dot{\vec r}_i|$,
$c$ the speed of light in vacuum, and
$\beta = \gamma = 1$ the nominal PPN factors.

% ??? Why there is G^2 in the expressions?! Well, its v_kepl^2/c^2...

We perform a numerical integration of orbits with the Bulirsch-Stoer algorithm.
It has an adaptive time step, with the relative precision set to $10^{-8}$.
The output time step is set to $0.2\,{\rm d}$,
but all times of observations are also integrated and output precisely.
The total number of free parameters is 38 (see Tab.~\ref{tab1}).
Observables are derived from coordinates, velocities,
radiative quantities, and distance.
Synthetic spectra, both relative and absolute, are computed during convergence
with Pyterpol \citep{Nemravova_2016A&A...594A..55N}
from BSTAR, OSTAR grids \citep{Lanz_2007ApJS..169...83L,Lanz_2003ApJS..146..417L}.

Observations and our model are compared by means of the $\chi^2$ metric:
\begin{eqnarray}
\chi^2 &=& w_{\rm rv}\chi^2_{\rm rv} + w_{\rm ttv}\chi^2_{\rm ttv} + w_{\rm vis}\chi^2_{\rm vis} + w_{\rm clo}\chi^2_{\rm clo} + w_{\rm t3}\chi^2_{\rm t3} \,+ \nonumber\\
&& +\, w_{\rm syn}\chi^2_{\rm syn} + w_{\rm sed}\chi^2_{\rm sed} + w_{\rm sky2}\chi^2_{\rm sky2} \,,
\end{eqnarray}
where subscripts denote the respective datasets:
radial velocities (RV),
eclipse/transit timing variations (TTV),
squared visibilities (VIS),
closure phases (CLO),
triple product amplitudes (T3),
synthetic spectra (SYN),
spectral-energy distribution (SED),
relative astrometry (SKY2).
Recent improvements of our model include also
fitting of metallicity,
subplex algorithm (i.e., simplex on sub-spaces; \citealt{Rowan_1990}),
precise computation of the Roche potential from volume-equivalent radius
for optional light curve computations (using $R(\Omega)$ integrated
as in \citealt{Leahy_2015ComAC...2....4L} and inverted to $\Omega(R)$).

\begin{figure}
\centering
\includegraphics[width=8cm]{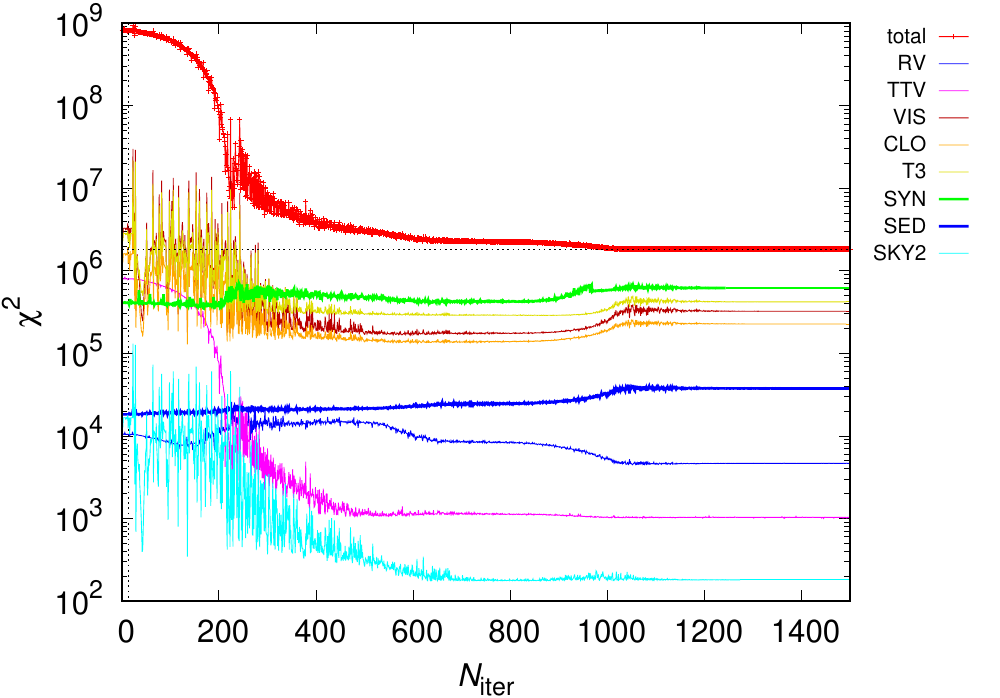}
\caption{
Example of convergence for the alternative model (see Tab.~\ref{tab1}).
All datasets (RV, TTV, VIS, CLO, T3, SYN, SED, SKY2) were taken into account,
with alternative weights
$w_{\rm rv} = 100$,
$w_{\rm ttv} = 1000$,
$w_{\rm vis} = 0.1$,
$w_{\rm clo} = 0.1$,
$w_{\rm syn} = 0.1$,
$w_{\rm sky2} = 1000.0$.
In this case, two stellar components (Ac2, Aa2) were 'forced' to be
on the main sequence, which results in `tension' (cf. increase for
several $\chi^2$ contributions after 800 iteration).
On the other hand, the nominal model does not exhibit such tension.
}
\label{chi2_iter}
\end{figure}

As a preparatory task, we computed models with parameters as
determined by the observation-specific models \citep{Mayer_2022arXiv220407045H}.
However, there were some caveats. First, orbital elements
in our N-body model (including periods $P_1$, $P_2$, $P_3$) are only osculating.
When a dynamical model is different, these elements must be
converged again.
Whether or not $\vec f_{\rm ppn}$ is included in the model,
also affects $\chi^2$;
in other words, to get accurate values of parameters,
it should be included.
Moreover, we had to `flip' the orientation
of the system several times to match some of the datasets (TTV, SKY2).
Only then we performed a convergence with the simplex algorithm
(\citealt{Nelder_Mead_1965}; e.g., Fig.~\ref{chi2_iter}).
Of course, this was performed multiple times using various initial conditions
to avoid a false convergence and to solve some remaining systematics (RV, SYN).

%%%%%%%%%%%%%%%%%%%%%%%%%%%%%%%%%%%%%%%%%%%%%%%%%%%%%%%%%%%%%%%%%%%%%%%%

\subsection{Survey of parameters}

In order to understand, how individual datasets constrain the model,
and to find a global minimum of $\chi^2$, we perfomed a survey of parameters.
We converged 81 different models.
Every convergence was initialized with a different combination of masses
$m_1$, $m_2$ (Ac1, Ac2), within in the range $15$ to $50\,M_{\rm S}$,
while $m_3$ (Aa1) was set to $m_{\rm sum}-m_1-m_2-m_4$;
nevertheless, {\em all\/} parameters were free during convergence.
The maximum number of iterations was set to 1000.
To speed up this computation, we used a set of predefined synthetic spectra.
We assumed unit weights, with the exception of $w_{\rm syn} = 0.1$.
The radiative parameters of the fourth spectrally undetected component (Aa2)
were constrained by \cite{Harmanec_1988BAICz..39..329H} parametric relations
$m_4(T_4)$, $\log g_4(T_4)$,
i.e., assuming it is a normal main-sequence object. The component is
too faint to significantly contribute to the total flux,
it is, however, not negligible from the dynamical point of view.

\begin{figure*}
\centering
\includegraphics[width=10.75cm]{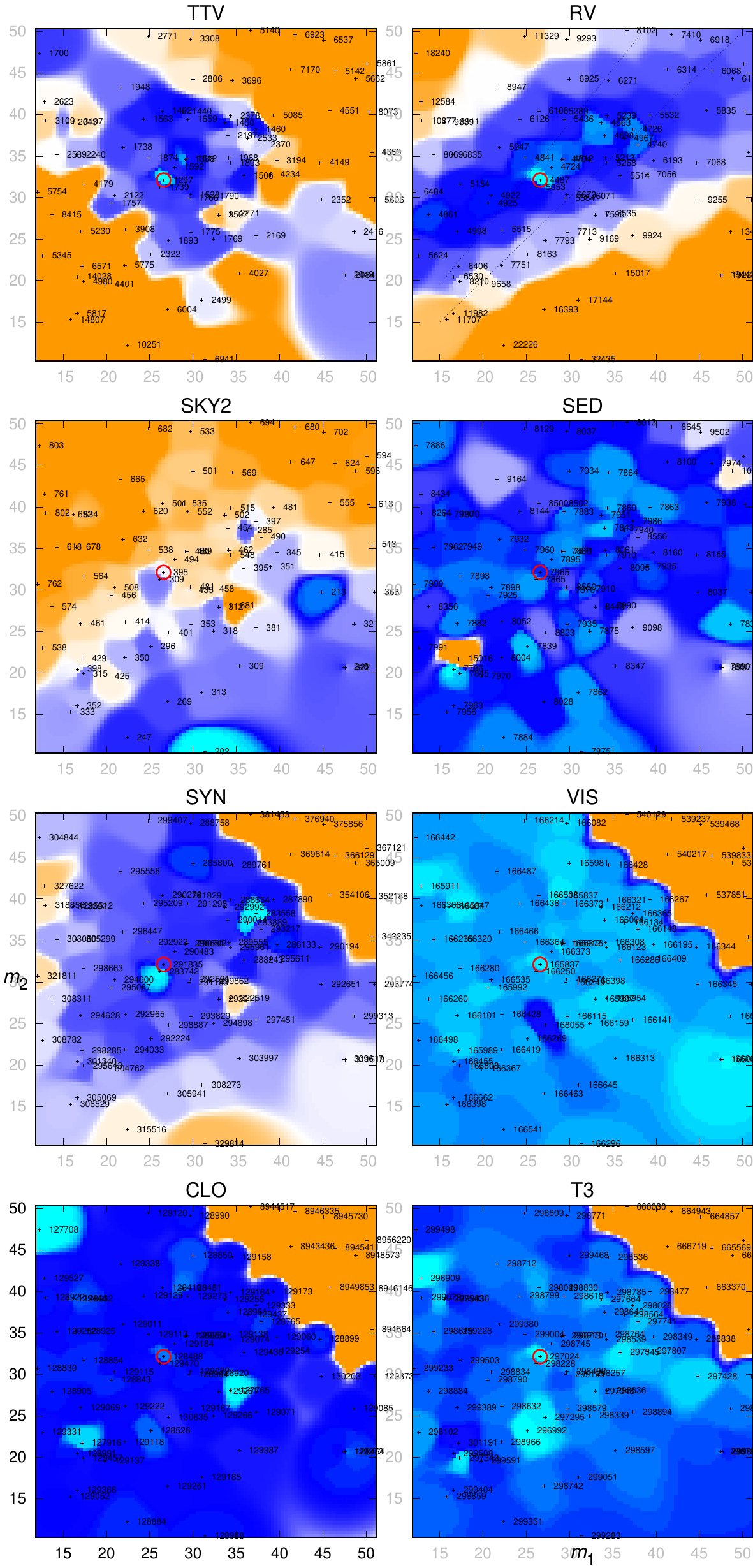}
\caption{
Contributions to $\chi^2$ for a set 81 best-fit models.
Individual contributions (datasets) are shown in the panels (from top left):
TTV,
RV,
SKY2,
SED,
SYN,
VIS,
CLO,
T3.
Every convergence was initialized with a different combination of masses
$m_1$, $m_2$ (Ac1, Ac2), within in the range $15$ to $50\,M_{\rm S}$,
while $m_3$ (Aa1) was set to $m_{\rm sum}-m_1-m_2-m_4$;
nevertheless, {\em all\/} parameters were free during convergence.
Axes correspond to the masses $m_1$, $m_2$,
colours to $\chi^2$ {\tiny (cf. tiny numbers)}, with adapted colour scales:
\color{cyan}cyan\color{black}\ the very best fit for dataset,
\color{blue}blue\color{black}\ good fits (${<}\,1.2\min\chi^2$),
\color{orange}orange\color{black}\ poor fits (${\ge}\,1.2\min\chi^2$).
The factor was 3.0 for TTV, RV, SKY2.
'Forbidden regions' can be clearly seen
(e.g., high $m_1$, $m_2$ especially due to CLO),
as well as correlations between parameters (TTV $-$, RV $+$).
The weighted very best fit for all datasets is denoted by
\color{red}red\color{black}\ circle.
}
\label{qzcar_fitting4_MINUSUVGRID_m1_m2_ALL}
\end{figure*}

The results are shown in Fig.~\ref{qzcar_fitting4_MINUSUVGRID_m1_m2_ALL}.
One can clearly see 'forbidden regions',
where no solution can be found ($\chi^2$ remains high).
For high $m_1$, $m_2$, this is especially due to the CLO and VIS datasets,
which strongly constrain the relative luminosities of (Ac1+Ac2), (Aa1+Aa2) components.
Moreover, there are strong correlations between parameters
--- for TTV negative, for RV positive.
This is very useful, because the best-fit model is just at the `intersection'.
Finally, the $\chi^2$ contribution due to the SED dataset seems to be too flat,
however, this is because all models were converged,
and the distance around 2800\,pc corresponds the SED.
It does not mean that the SED is unimportant!

%%%%%%%%%%%%%%%%%%%%%%%%%%%%%%%%%%%%%%%%%%%%%%%%%%%%%%%%%%%%%%%%%%%%%%%%

\subsection{Nominal model}\label{nominal}

\begin{figure}
\centering
\includegraphics[width=9cm]{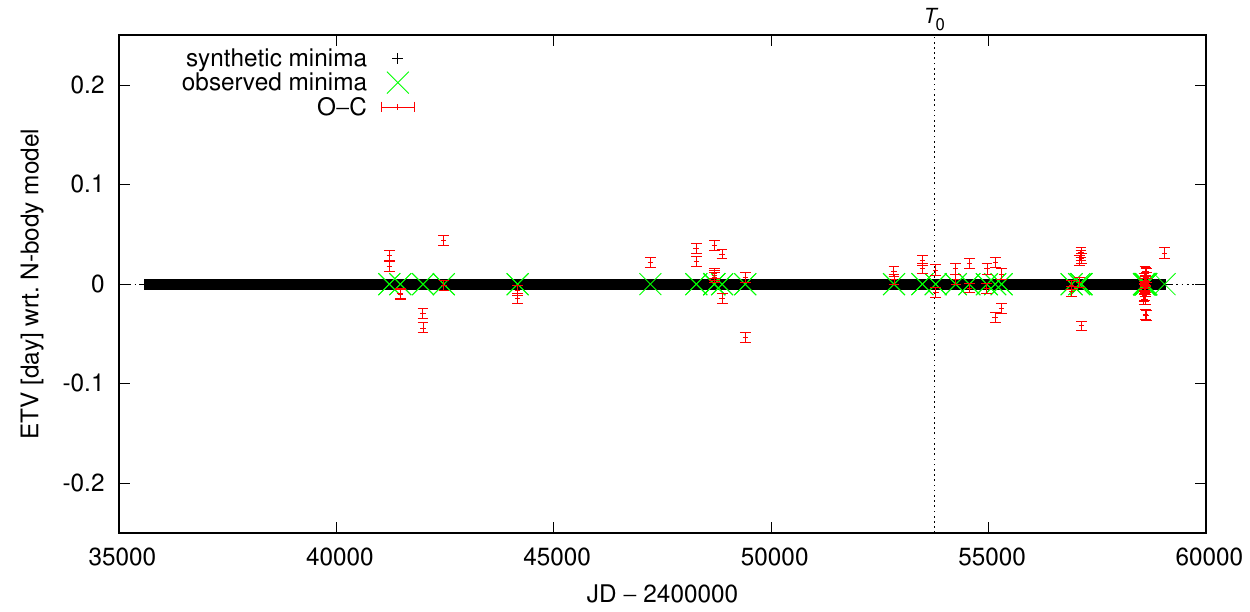}
\includegraphics[width=9cm]{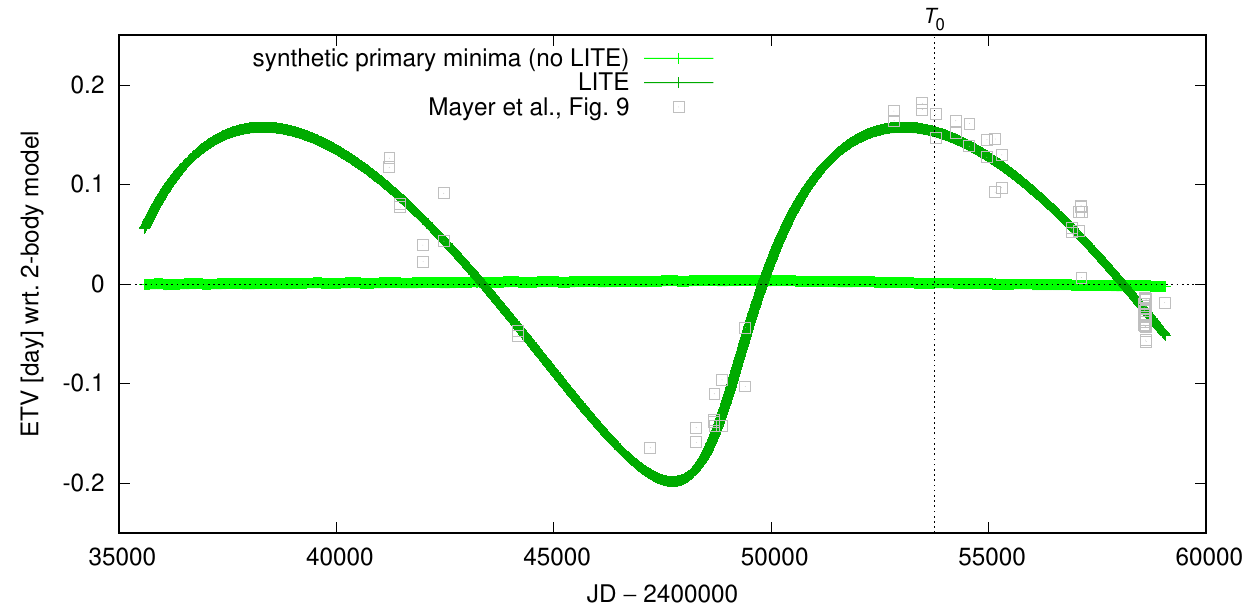}
\caption{
Eclipse timing variations (ETVs) for the nominal model (top; see Tab.~\ref{tab1}).
All synthetic minima (primary and secondary) are plotted on the $x$ axis (black),
together with the observed minima (\color{green}green\color{black})
and their difference on the $y$ axis (\color{red}red\color{black}).
In our N-body model, these ETVs are small,
because they are suppressed by the actual motion of the eclipsing binary (Ac1+Ac2)
about the centre of mass of all 4~components.
For comparison, a simplified two-body model from \cite{Mayer_2022arXiv220407045H},
exhibiting large ETVs, is plotted (bottom; \color{gray}gray\color{black}),
{\bf with the light-time effect extracted from our N-body model (\color{OliveGreen}olive\color{black}).}
}
\label{chi2_TTV}
\end{figure}

\begin{figure}
\centering
\includegraphics[width=9cm]{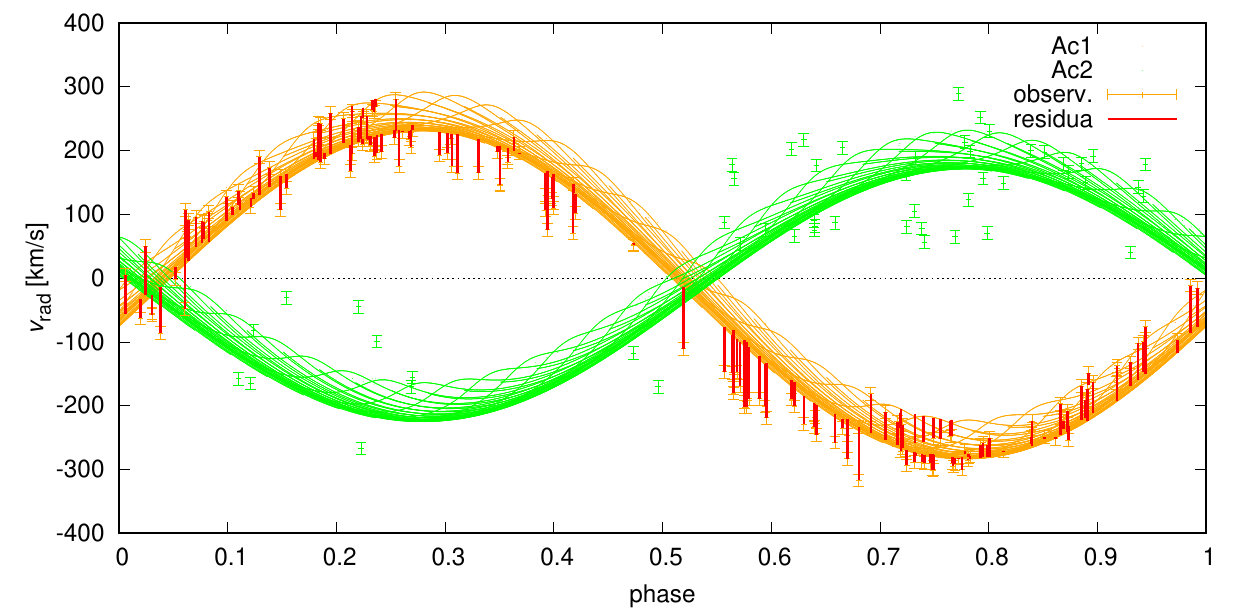}
\caption{
Radial velocities (RVs) of components
Ac1 (\color{orange}orange\color{black}),
Ac2 (\color{green}green\color{black}).
Observed values derived from spectra (\citealt{Mayer_2022arXiv220407045H}; error bars),
synthetic values directly derived from our N-body model (dots)
and residuals (\color{red}red\color{black}) are plotted.
Large residuals are present close to the eclipses (phases 0.0, 0.5),
because individual lines (RVs) cannot be reliably separated,
as well as for the component Ac2,
which has very broad lines (hence uncertain RVs).
}
\label{chi2_RV1}
\end{figure}

\begin{figure}
\centering
\includegraphics[width=9cm]{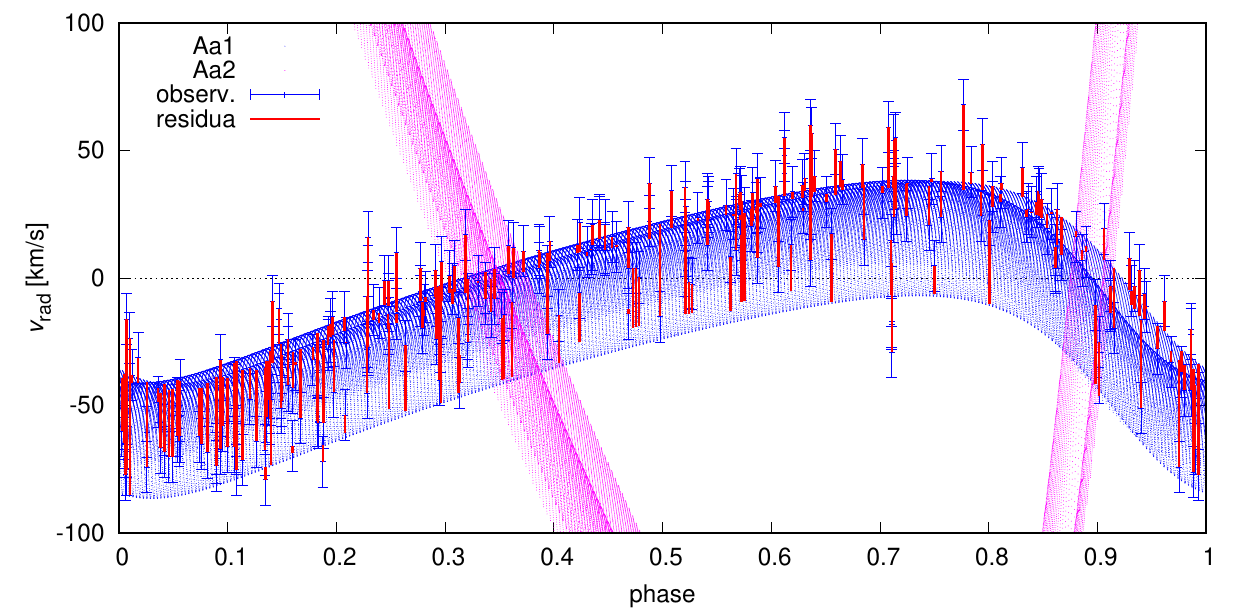}
\caption{
Same as Fig.~\ref{chi2_RV1} for components
Aa1 (\color{blue}blue\color{black}),
Aa2 (\color{magenta}magenta\color{black}).
The spectroscopic binary orbit (Aa1+Aa2) is highly eccentric.
The broad range of both synthetic and observed RVs
is due to its motion on the wide orbit ((Ac1+Ac2)+(Aa1+Aa2)).
}
\label{chi2_RV3}
\end{figure}

\begin{figure}
\centering
\includegraphics[width=9cm]{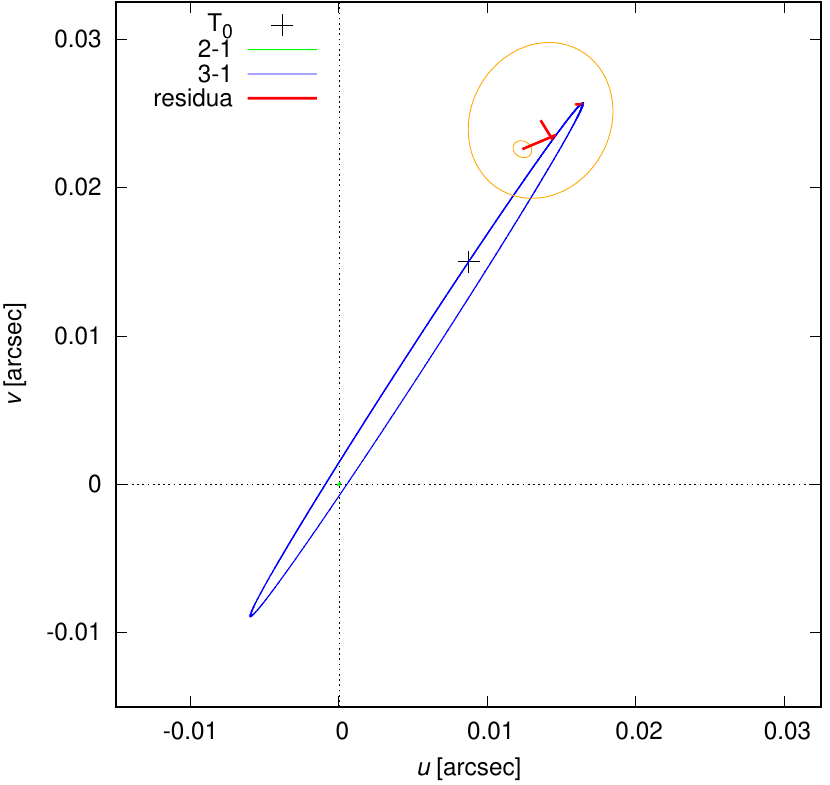}
\caption{
Astrometry derived from previous interferometric measurements
(see Tab.~\ref{tab:litpro}).
Positions of Aa1 components with respect to Ac1 are plotted
(\color{blue}blue\color{black}),
as well as Ac2 w.r.t. Ac1
(\color{green}green\color{black}; in the very centre).
Tiny oscillations visible as thickness of lines are
due to photocentre motions.
We used positive signs of $(u, v)$ for this plot.
The orbit is constrained not only by this astrometry,
but also by ETVs and RVs (see Figs.~\ref{chi2_TTV}, \ref{chi2_RV3}).
}
\label{chi2_SKY2}
\end{figure}

\begin{figure*}
\centering
\includegraphics[width=18.0cm]{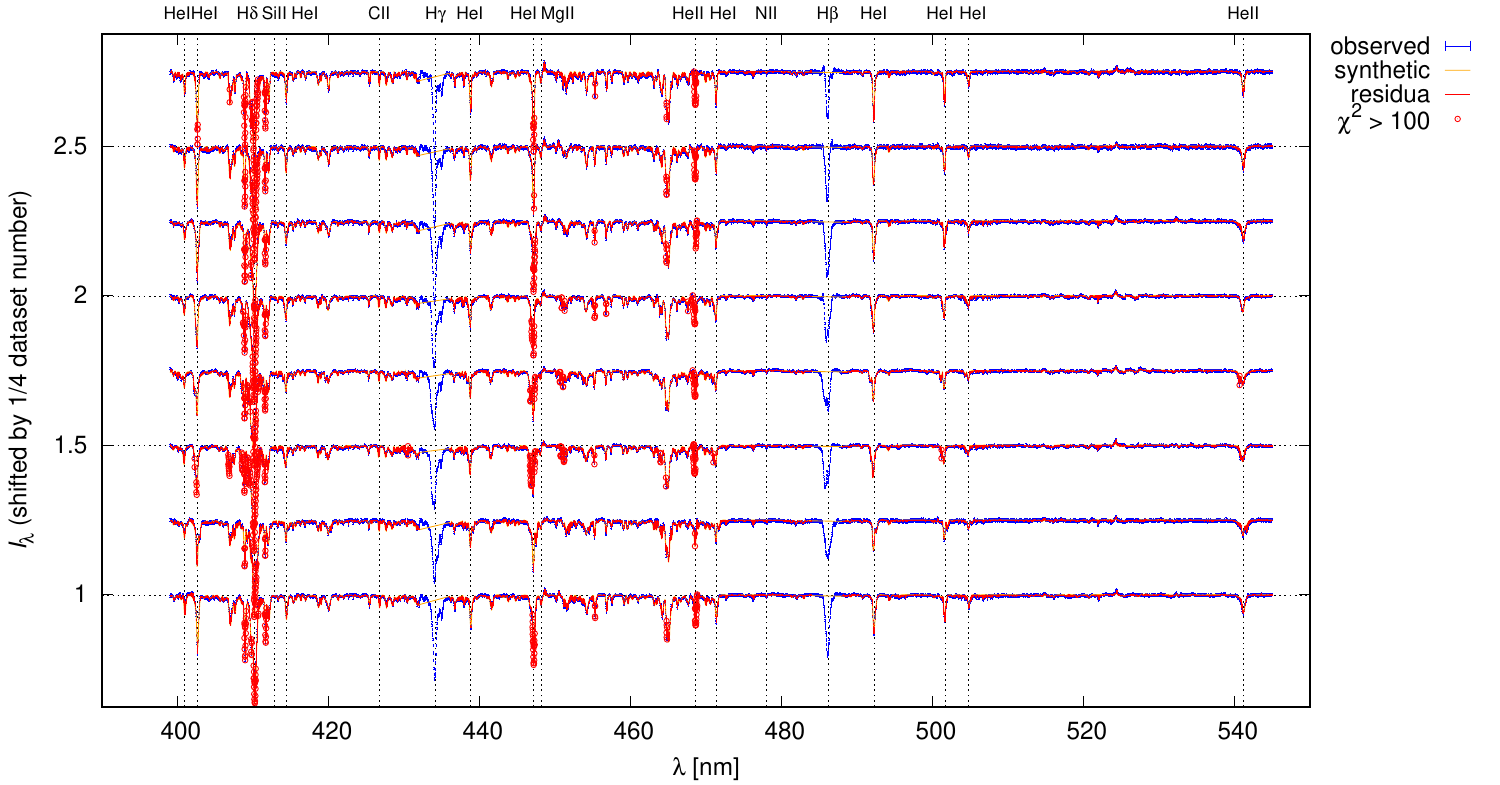}
\caption{
Comparison of the rectified observed FEROS spectra (\color{blue}blue\color{black})
with the synthetic spectra computed by our N-body model (\color{orange}orange\color{black}).
Residuals are also plotted (\color{red}red\color{black}).
They are substantial (see circles) for the $H\delta$ line,
where the continuum level is uncertain (between 405.1 and 412.6\,nm).
Nevertheless, most line features (including splits, blends, weak lines
close to the continuum) are present in both observed and synthetic spectra,
resulting in reliable RVs of individual stellar components
(especially Ac1, Aa1; cf.~Fig.~\ref{synthetic2}).
}
\label{chi2_SYN}
\end{figure*}

\begin{figure}
\centering
\includegraphics[width=9cm]{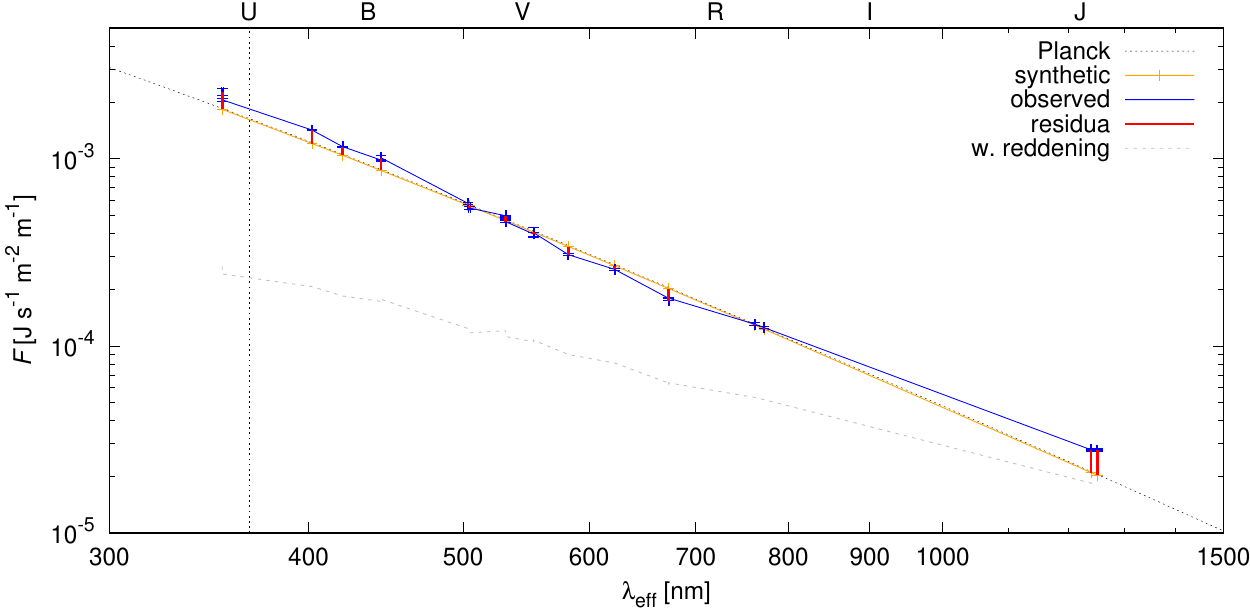}
\caption{
Comparison of the dereddened spectral-energy distribution
(SED; \color{blue}blue\color{black})
and the synthetic SED (\color{red}red\color{black}).
While the near-ultraviolet to near-infrared region
is reasonably described by our N-body model,
there is a substantial excess of the observed flux.
in the far-infrared region.
This is possibly related to non-thermal processes in stellar atmospheres
or presence of circumstellar matter,
which is not included in our N-body model.
}
\label{chi2_SED}
\end{figure}

\begin{figure*}
\centering
\includegraphics[width=18.0cm]{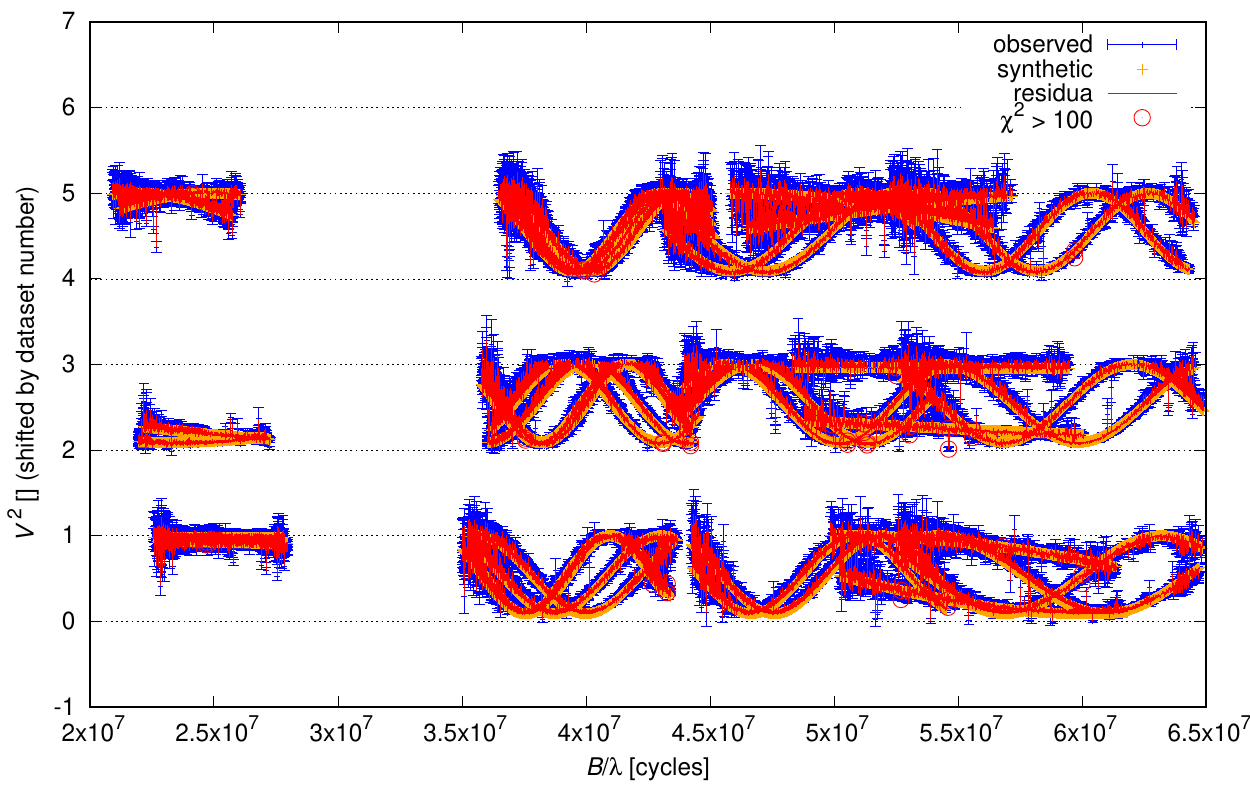}
\caption{
Squared visibilities $V^2$ for the VLTI/GRAVITY observations
(Mar 14th 2017, Apr 27th 2017; \color{blue}blue\color{black}),
synthetic $V^2$ (\color{orange}orange\color{black}),
and residuals (\color{red}red\color{black}).
The sinusoidal dependence on the baseline $B/\lambda$
essentially corresponds to a wide binary ((Ac1+Ac2)+(Aa1+Aa2)).
Disks of individual stellar components cannot be resolved.
(at $0.08\,{\rm mas}$, this would correspond a drop of $V^2$
by only 0.01 at the longest baseline).
}
\label{chi2_VIS}
\end{figure*}

\begin{figure}
\centering
\includegraphics[width=9cm]{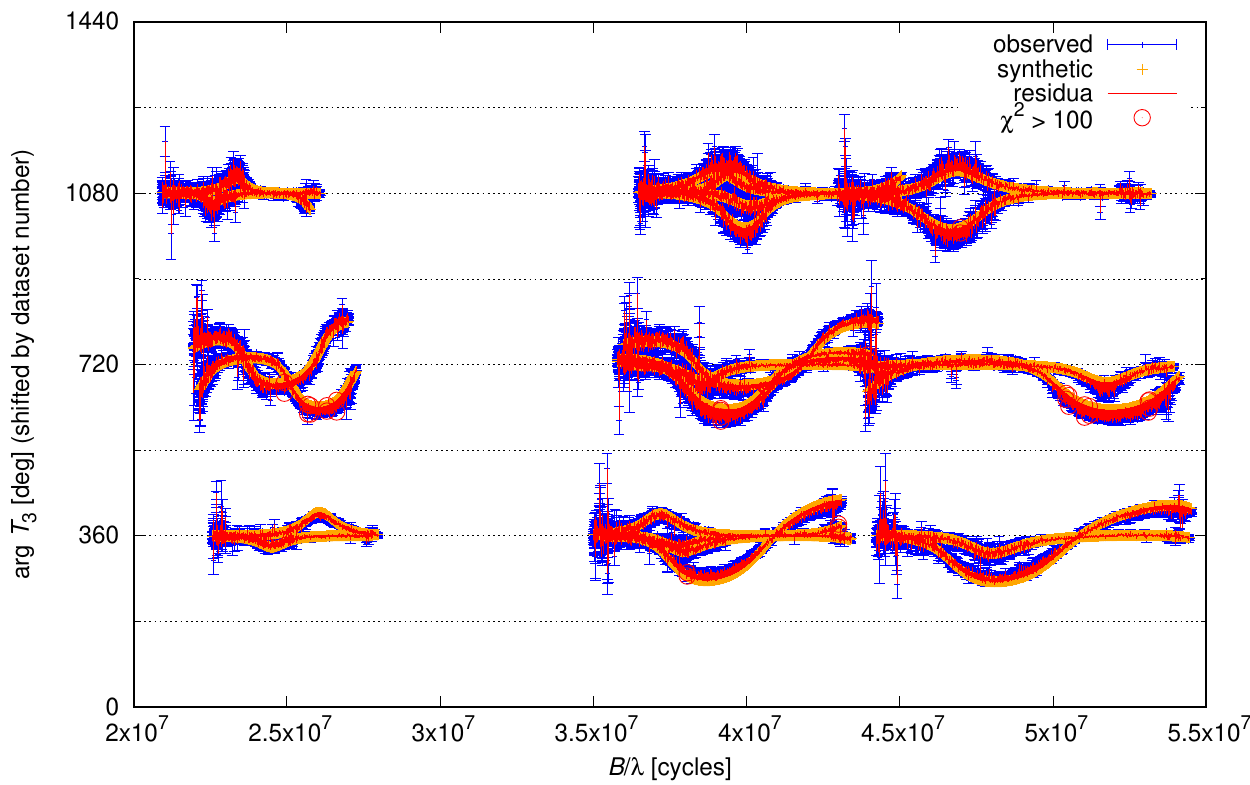}
\caption{
Same as Fig.~\ref{chi2_VIS} for the closure phase $\arg T_3$.
Again, the dependence on $B/\lambda$ is binary-like.
}
\label{chi2_CLO}
\end{figure}

\begin{figure}
\centering
\includegraphics[width=9cm]{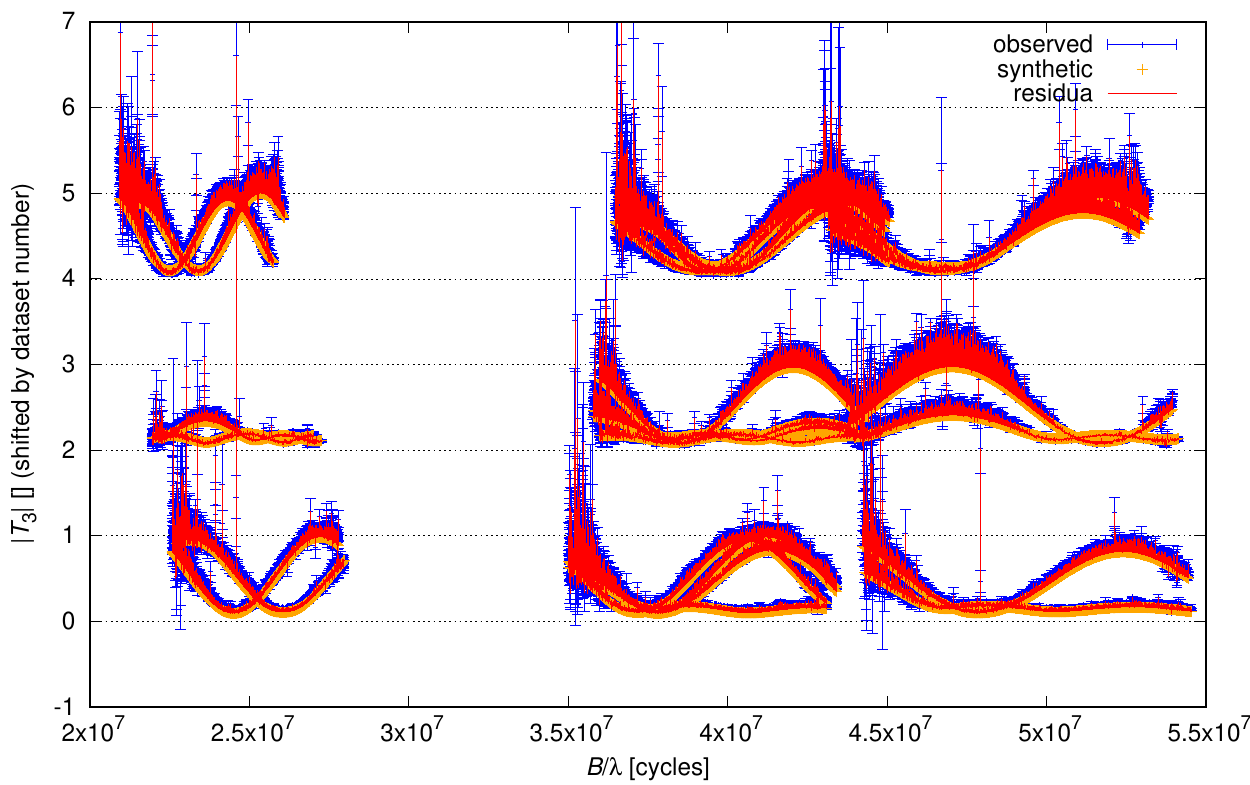}
\caption{
Same as Fig.~\ref{chi2_VIS} for the triple product amplitude $|T_3|$.
Observed values are often close to or even exceed~$1$,
which indicates a possible calibration problem at long baselines
($B/\lambda > 4\times 10^7$ cycles per baseline);
synthetic values are always ${<}\,1$.
Nevertheless, the positions of maxima and minima of $|T_3|(B/\lambda)$
are fitted perfectly.
}
\label{chi2_T3}
\end{figure}

The best-fit parameters of our nominal model are listed in Tab.~\ref{tab1}.
Observed and synthetic data are then compared in numerous
Figs.~\ref{chi2_TTV} to~\ref{chi2_T3}.

First, the eclipse-timing variations (Fig.~\ref{chi2_TTV})
exhibit much smaller amplitudes compared to \cite{Mayer_2022arXiv220407045H},
because they are suppressed by the actual motion of the
eclipsing binary (Ac1+Ac2). The respective $\chi^2_{\rm ttv}$
contribution is still larger than the number of data points $n_{\rm ttv}$,
because one has to fit other datasets at the same time (RV);
the result is a compromise.

Second, the radial velocities of the eclipsing binary Ac1+Ac2 (Fig.~\ref{chi2_RV1})
show some systematics related to Ac1 component
at the phases 0.0, 0.5 (i.e., primary and secondary eclipses),
when lines are blended and RVs should be close to zero.
Additional systematics are present elsewhere, e.g., between 0.6-0.7,
which cannot be avoided due to other RV measurements with relatively
low RV values. A reason might be hidded in a complex reduction
and rectification procedure; one possible solution would be to
prefer fitting of synthetic spectra, instead of deriving RVs.
(See also Section~\ref{oblateness}.)
Regarding Ac2 component, which is relatively faint
and fast-rotating, a number of measurements is offset.
Because these substantially contribute to $\chi^2_{\rm rv}$,
we decided to remove RVs of Ac2 in the alternative model (see below).

Third, the RVs of Aa1 component (Fig.~\ref{chi2_RV3}) seem to
be reliable, because it is relatively bright and slow-rotating.
Yet, both Aa1 and Ac1 RVs are clearly affected by the wide orbit ((Ac1+Ac2)+(Aa1+Aa2)).
This was detected in \cite{Mayer_2022arXiv220407045H} as variable $\gamma$ velocities.
The fourth component Aa2 is too faint, but its predicted maximum RV
should be of the order of ${\pm}\,300\,{\rm km}\,{\rm s}^{-1}$.

Astrometry of the wide orbit (Ac+Aa `binary'; Fig.~\ref{chi2_SKY2})
provides a fundamental angular scale, which together with ETVs
and RVs constrains the fundamental parameters. Even though there is
some `tension' between periods from RVs and ETVs \citep{Mayer_2022arXiv220407045H},
we regard the resulting period $P_3 = (14722\pm 50)\,{\rm d}$
as a reasonable compromise.

The blue part (399--544\,nm) of the eight rectified FEROS spectra
(with high signal-to-noise ratio)
were also fitted (Fig.~\ref{chi2_SYN}).
The $\chi^2_{\rm syn}$ contribution is still substantial,
mostly because the level of continuum was not always correctly determined.
H$\beta$, H$\gamma$ lines were removed from the fitting procedure
due to strong emission features;
impossible to be fitted by our model,
which does not include any optically-thin circumstellar matter (CSM).
We can only fit weak emission lines (e.g., \ion{Fe}{III}),
present in synthetic spectra of stellar atmospheres.
On the other hand, H$\delta$ does have features (asymmetries),
which are reasonably fitted, although the depths of synthetic hydrogen lines
are not matched precisely, which is --again-- related to the level
of surrounding continuum (for example, between 405.1 to 412.6\,nm
there is no continuum, as verified by synthetic spectra).
Nevertheless, the depths as well as features of
\ion{He}{I},
\ion{C}{II},
\ion{O}{II},
\ion{Mg}{II}
lines are fitted much better.
Overall, the determination of the respective RVs is very precise
(especially for Ac1, Ac2 components),
because we fit {\em all} lines at the same time
(a.k.a. cross-correlation).
Our model also naturally couples all spectra (by the orbits),
which avoids line blends.

Regarding the SED (Fig.~\ref{chi2_SED}), the fit is acceptable,
within $\pm 0.2\,{\rm mag}$ in the NUV--NIR range,
which is likely related to the uncertainties of dereddening
(described in Sec.~\ref{sed}).
Moreover, 0.15\,mag is the depth of minima.
The FIR range is not fitted, without any CSM,
which would indeed increase the respective $\chi^2_{\rm sed}$ contribution.
Nevertheless, the SED
(together with the masses, radial velocities, angular scale, \dots)
determines the distance around 2800\,pc.

A comparison of interferometric observations and our model
is straightforward for the squared visibilities~$V^2$
(Fig.~\ref{chi2_VIS}). 
The sinusoidal dependence on the baseline $B/\lambda$
essentially corresponds to a wide binary ((Ac1+Ac2)+(Aa1+Aa2)).
Scatter of observed $V^2$ is related to the Poisson noise;
our synthetic $V^2$ are smooth.
Because $V^2$ values can be very close to $1$ (for a binary),
it is important to include observed values
which are slightly~${>}\,1$, as well as~${<}\,1$,
because only then the mean value is close to~$1$.
The corresponding precision of astrometry,
when all VLTI/GRAVITY measurements are `compressed' to $(u,v)$ coordinates,
is of the order of $0.010\,{\rm mas} = 10\,\mu{\rm as}$.

The same is true for the closure phases $\arg T_3$ (Fig.~\ref{chi2_CLO}).
Similarly as in \cite{Sanchez_2017ApJ...845...57S}
the dependence on $B/\lambda$ is binary-like.

On contrary, the triple product amplitude (Fig.~\ref{chi2_T3})
exhibits some systematics, especially at long baselines
($B/\lambda > 4\times 10^7$ cycles per baseline);
observed values are often close to or even exceed~$1$,
which seems to be a calibration problem.
Nevertheless, the positions of maxima and minima of $|T_3|$
as a function of $B/\lambda$ are fitted perfectly,
hence we retained the dataset in the nominal model
(and we removed it in the alternative model; see Sec.~\ref{alternative}).

%%%%%%%%%%%%%%%%%%%%%%%%%%%%%%%%%%%%%%%%%%%%%%%%%%%%%%%%%%%%%%%%%%%%%%%%

\begin{figure}
\centering
\includegraphics[width=9cm]{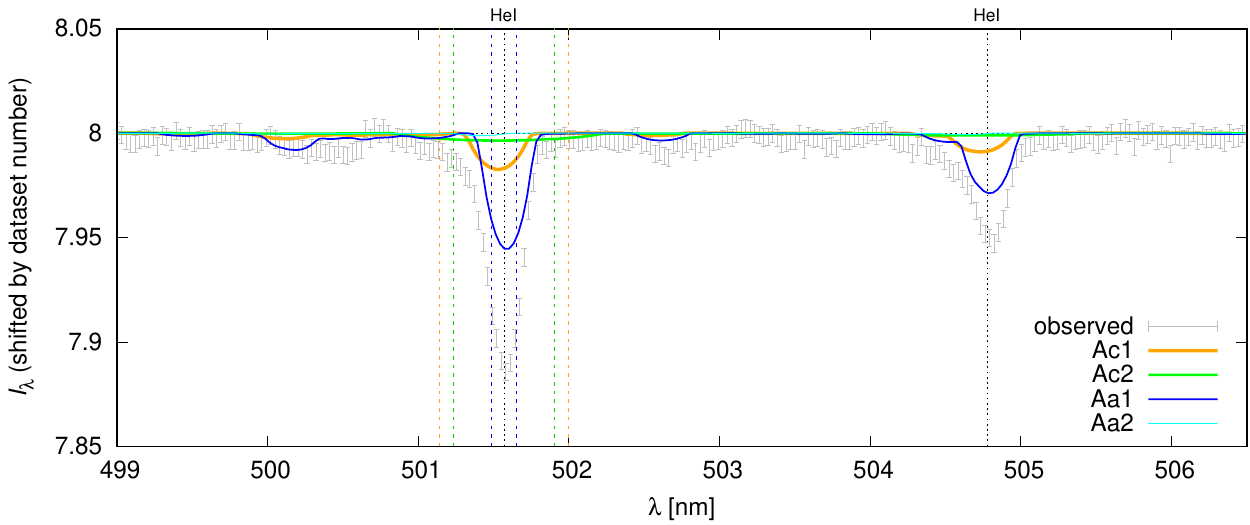}
\caption{
Synthetic spectra of individual components (Ac1, Ac2, Aa1, Aa2)
weighted by their luminosities,
so that one can see their contributions to the total spectrum.
For comparison, the observed spectrum is plotted (error bars).
The region of \ion{He}{I} 5016 and 5047 lines is plotted.
Ac2 component is fast-rotating and has broad lines,
Aa2 component is not luminous and its contribution is negligible.
}
\label{synthetic2}
\end{figure}

%%%%%%%%%%%%%%%%%%%%%%%%%%%%%%%%%%%%%%%%%%%%%%%%%%%%%%%%%%%%%%%%%%%%%%%%

\subsection{Alternative model}\label{alternative}

Alternative models are useful to determine the true uncertainties
of parameters. With this goal in mind, we modified a number of
things, in particular, RVs of Ac2 component were not used,
because they did not seem reliable. Instead, Ac2 component
was again constrained according to \cite{Harmanec_1988BAICz..39..329H}.
In case of RVs of Ac1, we removed outliers identified in the
nominal model.

Similarly, $|T_3|$ was not used, because it exhibited systematics
at long baselines, while $V^2$ at the same baselines was fitted precisely.
Outliers were also removed from the VIS, CLO, and T3 datasets,
as they often occur at the beginning of the `burst',
so that we assume they were caused by temporarily erroneous setup.

The weights were set as `extreme', to enforce fitting of datasets
with low number of points:
$w_{\rm rv} = 100$,
$w_{\rm ttv} = 1000$,
$w_{\rm vis} = 0.1$,
$w_{\rm clo} = 0.1$,
$w_{\rm syn} = 0.1$,
$w_{\rm sky2} = 1000.0$.
On contrary, we relaxed the SED constraint by using $w_{\rm sed} = 1$.

The results are summarized in Tab.~\ref{tab1} and Figures were
moved to Appendix~\ref{appendix}.
Major differences between the models can summarized as follows.
Using \cite{Harmanec_1988BAICz..39..329H} means forcing Ac2 component
to be on the main sequence. Since we needed a substantial mass ratio
to explain RVs of Ac1, the temperature of Ac2 increases
(and its radius decreases).
Putting less weight on the SED means that the distance is allowed
to decrease (down to 2500\,pc),
at the expense of systematic differences in the V--NIR region.
ETVs are more centered with respect to zero $O-C$ value,
including a group of measurements at ${\rm JD} \simeq 2458570$.
RVs and astrometry of Aa1 indicate that the eccentricity
of the wide orbit is still uncertain ($\log e_3 = -0.28$ vs. $-0.54$).
A comparison of observed and synthetic spectra shows
somewhat larger systematics for \ion{He}{I} lines.
For interferometric quantities ($V^2$, $\arg T_3$),
we detect only minor shifts, in $B/\lambda$,
and offsets in amplitudes of the closure phase.

Even though the alternative model exhibits larger total $\chi^2$,
it is a viable alternative and we can use it to estimate
uncertainties (see Tab.~\ref{tab1}; last column).

%%%%%%%%%%%%%%%%%%%%%%%%%%%%%%%%%%%%%%%%%%%%%%%%%%%%%%%%%%%%%%%%%%%%%%%%

\subsection{A note on oblateness}\label{oblateness}

Oblateness of components may also contribute to the precession
of orbits. It can be characterized by
the Love number~$k_2$,
the rotation period~$P$, and
the body radius~$R$
(as in \citealt{Fabrycky_2010exop.book..217F}),
or alternatively by a series of multipoles $C_{\ell,m}$, $S_{\ell,m}$
if the respective shape is more complex
(as in \citealt{Broz_2021A&A...653A..56B}).
For binary stars, $k_2$ is primarily determined by the density profiles
and the Roche potential
\citep{Claret_2004A&A...424..919C}.
For `soft' bodies, with the polytropic index $n = 3$ to $4$,
the expected value is of the order $10^{-2}$ to $10^{-3}$
\citep{Yip_2017MNRAS.472.4965Y};
it's even less for evolved stars with extended envelopes.%
\footnote{For comparison, an incompressible body has $k_2 = 0.75$ (exact),
and the Earth $0.295$ \citep{Lainey_2016CeMDA.126..145L}.}

According to our computations, the oblateness with $k_2 = 10^{-3}$
increases $\chi^2_{\rm ttv}$ up to 5660, but it can be compensated
by slightly adjusting the period~$P_1$.
Only if $k_2$ reaches ${\sim}\,0.003$ for Ac1 component,
which is probably the upper limit, because $\log g_1 \lesssim 3.5$,
it would increase the precession rate $\dot\omega_1$,
hence $\chi^2_{\rm ttv}$ as well as $\chi^2_{\rm rv}$.
Interestingly, the difference appears to be of the same order
as the systematics of RVs mentioned in Section~\ref{nominal}.
Nevertheless, the precession is not so substantial,
observed RV curves do not exhibit a clear temporal trend,
but rather a scatter \citep{Mayer_2022arXiv220407045H},
and thus all the parameters (except $P_1$) remain the same,
within uncertainties.

%%%%%%%%%%%%%%%%%%%%%%%%%%%%%%%%%%%%%%%%%%%%%%%%%%%%%%%%%%%%%%%%%%%%%%%%

\section{Discussion}\label{sec:discussion}

Compared to previous works on QZ~Car
\citep{Walker_2017MNRAS.470.2007W,Blackford_2020JAVSO..48....3B},
our new model indicates a higher total mass ($137$ vs. $112\,M_{\rm S}$)
and a different distribution of masses within the binaries Aa1+Aa2, Ac1+Ac2,
preferring the mass ratios Aa1/Aa2 ${\simeq}8$,
and Ac1/Ac2 close to 1, respectively.

% Walker etal. (2017) -- spectroscopy and photometry
% masses m1 = 43  , m2 = 19  , m3 = 30  , m4 = 20 M_S   +- 
% vs.    m1 = 26.1, m2 = 32.2, m3 = 70.2, m4 = 8.8 M_S
% total mass 112 vs. 137.3 M_S
% distance (2700+-300) vs. (2800+-100) pc

Given the photo--spectro--interferometric distance of $(2800\pm100)\,{\rm pc}$
for the nominal model, QZ~Car is likely related to the Carina Nebula (NGC\,3372).
However, when cluster distances were determined from the Gaia early data release~3 parallaxes
\citep{Shull_2021ApJ...914...18S,Goppl_2022A&A...660A..11G},
they turned out to be systematically smaller,
possibly due to anomalous extinction,
with $R_V \equiv A_V/E(B-V) = 3.2$ to $4.0$.
For Collinder 228, the median distance is 2470\,pc
and the angular size about $14'$,
which corresponds to the tangential size of only $10\,{\rm pc}$.
(The radial size is not well constrained by parallaxes.)
This revised distance is in accord with our alternative model of QZ~Car.

Massive multiple stars in this region may shed light on
possible progenitors of the $\eta$~Carin\ae\ event \citep{Weigelt_2016A&A...594A.106W},
which was suggested to be caused by mass transfer in triple systems,
eventually leading to an explosive merger \citep{Smith_2018MNRAS.480.1466S}.
Notably, QZ Car quadruple system includes the eclipsing mass-transferring
binary Ac1$\rightarrow$Ac2, which will evolve substantially.
According to our model, the mass ratio Ac2/Ac1 is relatively close to~1;
it means that we observe this binary almost at its closest distance.
If the final mass of Ac1 will decrease to, e.g., $5\,M_{\rm S}$,
the angular momentum conservation implies the final separation
(as well as the envelope extent of Ac1; \citealt{Paxton_2015ApJS..220...15P}),
of the order of $1\,000\,R_{\rm S}$,
i.e., substantially less than the pericentre of (Aa1+Aa2)+(Ac1+Ac2) binary,
$a_3(1-e_3) = 6\,300\,R_{\rm S}$.
This does not lead to an imminent instability.%
\footnote{Interestingly, Aa1 component seems to be exceedengly massive
(up to $70\,M_{\rm S}$), especially compared to Aa2 component (less than $10\,M_{\rm S}$).
This may be also related to a (putative) past mass transfer Aa2$\,\rightarrow\,$Aa1.}
It is still unclear, what has been (and shall be) the role of loosely bound companions,
Ab, Ad, B, C, D,
imaged around QZ~Car \citep{Rainot_2020A&A...640A..15R},
because we still lack their proper motions.

\section{Conclusions}\label{sec:conclusions}

Using \cite{Mayer_2022arXiv220407045H} observation-specific models as initial conditions for convergence,
we constructed a robust N-body model of the QZ~Car quadruple system,
fitting all three orbits and stellar--radiative parameters at the same time.
Because we used all types of observations,
which are in a sense orthogonal,
the model is well-behaved and does not exhibit strong correlations (`drifts').
Independent constraints for individual parameters were only used
when needed (e.g., for the luminosity--radius of Aa2 component,
which is too faint to be directly observable in the flux).

Our preferred model is the nominal one (see Tab.~\ref{tab1});
the best-fit masses are
$m_1 = 26.1\,M_{\rm S}$,
$m_2 = 32.3\,M_{\rm S}$,
$m_3 = 70.3\,M_{\rm S}$,
$m_4 = 8.8\,M_{\rm S}$,
with uncertainties of the order of $2\,M_{\rm S}$,
and the distance
$d = (2800\pm 100)\,{\rm pc}$.
The alternative model,
with Ac2 component `forced' to be on the main sequence,
exhibits some systematic differences,
especially for the SED,
which can be attributed to anomalous extinction with $R_V \sim 3.4$.
We used it to determine realistic uncertainties of parameters.

There are still some poorly constrained parameters,
in particular, the mutual inclinations of the orbits,
because the longitude of nodes~$\Omega_1$ is not constrained by eclipses.
For simplicity, we assumed a co-planarity,
but observations with the VLTI/GRAVITY instrument covering both short periods
($P_1$, $P_2$) could be used to constrain it,
because astrometric uncertainties related to Aa1 component
are of the same order as the photocentre motions related to
the eclipsing binary Ac1+Ac2.

As a future work, we suggest to include also the observed light curve
and strong emission lines (especially, H$\alpha$, H$\beta$) in the model.
However, this is generally difficult, given a possible presence
of both optically-thick and optically-thin circumstellar matter (CSM).
An appropriate treatment of the radiation transfer in the CSM
in needed for this purpose (e.g., \citealt{Broz_2021A&A...645A..51B}).
Moreover, a presence of oscillatory signals in the TESS light curve
also requires substantial model improvements (e.g., \citealt{Conroy_2020ApJS..250...34C}).

% relax co-planarity, i_1 = i_2 = i_3?
% fit also metallicities of components?
% use minima duration? (well, not clear minimum)
% J_2 of two point masses
% check Roche lobe filling factor of Ac1! (important constraint)
% 1 .. Ac1 .. O8III
% 3 .. Aa1 .. O9.7Ib
% Halpha emission Aa1 (wind)
% mass transfer Ac1->Ac2
% Ac2 should be surrounded by a disk (thick or thin)?
% Phoebe: R1 = 19.5 R_S, R2 = 10.3 R_S
% How the lightcurve is affected by a disk (cf. beta Lyr A)?!

%%%%%%%%%%%%%%%%%%%%%%%%%%%%%%%%%%%%%%%%%%%%%%%%%%%%%%%%%%%%%%%%%%%%%%%%

\begin{acknowledgements}
M.B., P.H. and M.W. were supported by the Czech Science Foundation grant 19-01995S.
We thank an anonymous referee for comments.
We thank Michael Shull for a~discussion about anomalous extinction.
\end{acknowledgements}

%%%%%%%%%%%%%%%%%%%%%%%%%%%%%%%%%%%%%%%%%%%%%%%%%%%%%%%%%%%%%%%%%%%%%%%%

\bibliographystyle{aa}
\bibliography{references}

%%%%%%%%%%%%%%%%%%%%%%%%%%%%%%%%%%%%%%%%%%%%%%%%%%%%%%%%%%%%%%%%%%%%%%%%

\begin{table*}
\caption{Best-fit parameters for the nominal (Sec.~\ref{nominal})
and alternative (Sec.~\ref{alternative}) models.
}
\label{tab1}
\centering
\begin{tabular}{llrrlr}
& & {\bf nominal} & alternative \\
comp. & var. & val. & val. & unit & $\sigma$ \\
\hline
\vrule height10pt width0pt
Ac1                 & $m_1          $ & $     25.5       $ & $31.4       $ & $M_{\rm S}$              & $ 2                 $ \\
Ac2                 & $m_2          $ & $     33.2       $ & $34.8^\dag  $ & $M_{\rm S}$              & $ 2                 $ \\
Aa1                 & $m_3          $ & $     69.8       $ & $55.2       $ & $M_{\rm S}$              & $ 5                 $ \\
Aa2                 & $m_4          $ & $     8.85^\dag  $ & $10.3^\dag  $ & $M_{\rm S}$              & $ 2                 $ \\
Ac1+Ac2             & $P_1          $ & $     5.998599   $ & $5.998601   $ & day                      & $ 0.000002          $ \\
                    & $\log e_1     $ & $     -2.876     $ & $-3.492     $ & 1                        & $ 1                 $ \\
                    & $i_1          $ & ${\bf +93.0     }$ & $-93.0      $ & deg                      & $ 1.0               $ \\
                    & $\Omega_1     $ & ${\bf 326.5     }$ & $146.0      $ & deg                      & $ 1.0               $ \\
                    & $\varpi_1     $ & $     343.4      $ & $15.9       $ & deg                      & $ 10.0              $ \\
                    & $\lambda_1    $ & $     266.4      $ & $265.6      $ & deg                      & $ 1.0               $ \\
Aa1+Aa2             & $P_2          $ & $     20.734367  $ & $20.734963  $ & day                      & $ 0.000050          $ \\
                    & $\log e_2     $ & $     -0.382     $ & $-0.451     $ & 1                        & $ 0.1               $ \\
                    & $i_2          $ & ${\bf +93.0     }$ & $-93.0      $ & deg                      & $ 1.0               $ \\
                    & $\Omega_2     $ & ${\bf 326.0     }$ & $146.0      $ & deg                      & $ 1.0               $ \\
                    & $\varpi_2     $ & $     83.8       $ & $79.2       $ & deg                      & $ 10.0              $ \\
                    & $\lambda_2    $ & $     306.3      $ & $305.3      $ & deg                      & $ 1.0               $ \\
Ac+Aa               & $P_3          $ & $     14728      $ & $14645      $ & day                      & $ 100               $ \\
                    & $\log e_3     $ & $     -0.268     $ & $-0.516     $ & 1                        & $ 0.1               $ \\
                    & $i_3          $ & ${\bf +92.1     }$ & $-93.2      $ & deg                      & $ 1.0               $ \\
                    & $\Omega_3     $ & ${\bf 326.7     }$ & $146.8      $ & deg                      & $ 1.0               $ \\
                    & $\varpi_3     $ & $     295.7      $ & $296.5      $ & deg                      & $ 10.0              $ \\
                    & $\lambda_3    $ & $     53.3       $ & $50.8       $ & deg                      & $ 1.0               $ \\
Ac1                 & $T_1          $ & $     29687      $ & $31894      $ & K                        & $ 1000              $ \\
Ac2                 & $T_2          $ & $     32979      $ & $43349      $ & K                        & $ 1000              $ \\
Aa1                 & $T_3          $ & $     29564      $ & $27705      $ & K                        & $ 1000              $ \\
Aa2                 & $T_4          $ & $     23446      $ & $25329      $ & K                        & $ 1000              $ \\
Ac1                 & $\log g_1     $ & $     3.40       $ & $3.69       $ & cgs                      & $ 0.1               $ \\
Ac2                 & $\log g_2     $ & $     3.80       $ & $3.92^\dag  $ & cgs                      & $ 0.1               $ \\
Aa1                 & $\log g_3     $ & $     3.37       $ & $3.35       $ & cgs                      & $ 0.1               $ \\
Aa2                 & $\log g_4     $ & $     4.10^\dag  $ & $4.10^\dag  $ & cgs                      & $ 0.1               $ \\
Ac1                 & $v_{\rm rot1} $ & $     121        $ & $129        $ & ${\rm km}\,{\rm s}^{-1}$ & $ 10                $ \\
Ac2                 & $v_{\rm rot2} $ & $     410        $ & $310^{\rm f}$ & ${\rm km}\,{\rm s}^{-1}$ & $ 100               $ \\
Aa1                 & $v_{\rm rot3} $ & $     109        $ & $117        $ & ${\rm km}\,{\rm s}^{-1}$ & $ 10                $ \\
Aa2                 & $v_{\rm rot4} $ & $     100^{\rm f}$ & $100^{\rm f}$ & ${\rm km}\,{\rm s}^{-1}$ & $ \hbox{--}         $ \\
                    & $\gamma       $ & $     -10.0      $ & $-13.6      $ & ${\rm km}\,{\rm s}^{-1}$ & $ 5                 $ \\
                    & $d_{\rm pc}   $ & $     2836       $ & $2452       $ & pc                       & $ 100               $ \\
\hline
\vrule width 0pt height 9pt
& $n_{\rm rv}        $ & $ 491     $ & $ 491     $ \\
& $n_{\rm ttv}       $ & $ 64      $ & $ 64      $ \\
& $n_{\rm vis}       $ & $ 54394   $ & $ 54394   $ \\
& $n_{\rm clo}       $ & $ 41367   $ & $ 41367   $ \\
& $n_{\rm t3}        $ & $ 41367   $ & --          \\
& $n_{\rm syn}       $ & $ 36800   $ & $ 36800   $ \\
& $n_{\rm sed}       $ & $ 32      $ & $ 32      $ \\
& $n_{\rm sky2}      $ & $ 12      $ & $ 12      $ \\
\hline
\vrule width 0pt height 9pt
& $\chi^2_{\rm rv}   $ & $ 4210    $ & $ 4142    $ \\
& $\chi^2_{\rm ttv}  $ & $ 1133    $ & $ 944     $ \\
& $\chi^2_{\rm vis}  $ & $ 165821  $ & $ 176818  $ \\
& $\chi^2_{\rm clo}  $ & $ 128730  $ & $ 125861  $ \\
& $\chi^2_{\rm t3}   $ & $ 296025  $ & --          \\
& $\chi^2_{\rm syn}  $ & $ 269954  $ & $ 291833  $ \\
& $\chi^2_{\rm sed}  $ & $ 7590    $ & $ 19144   $ \\
& $\chi^2_{\rm sky2} $ & $ 349     $ & $ 59.4    $ \\
& $\chi^2            $ & $ 630856  $ & $ 1759054^{\rm w} $ \\
\hline
\end{tabular}
\tablefoot{
Notation in this table is:
1 .. Ac1,
2 .. Ac2,
3 .. Aa1,
4 .. Aa2.
Orbital elements are osculating
for the epoch $T_0 = 2453738.819100$, where
$m_1$ denotes the mass of body~1,
and so on for bodies 2, 3, 4;
$P_1$ the orbital period of the orbit (1+2),
$\log e_1$ logarithm of eccentricity,
$i_1$ inclination,
$\Omega_1$ longitude of node,
$\varpi_1$ longitude of pericentre,
$\lambda_1$ true longitude,
and so on for orbits (3+4), (1+2)+(3+4);
$T_1$ effective temperature of body~1,
$\log g_1$ logarithm of gravity,
$v_{\rm rot1}$ rotational velocity,
$\gamma$ systemic velocity,
$d_{\rm pc}$ distance,
$n$~denote numbers of observations (RV, TTV, VIS, CLO, T3, SYN, SED, SKY2),
$\chi^2$ is simply unreduced $\chi^2$.
The nominal and alternative models use different weights (cf.~$^{\rm w}$).
Parameters denoted as $^\dag$ were derived from~$T$, according to \cite{Harmanec_1988BAICz..39..329H},
and $^{\rm f}$ were fixed.
The angular orbital elements are expressed in the standard stellar reference frame;
(Ac1+Ac2) eclipsing binary is located in the centre.
}
\end{table*}

\begin{table}
\caption{Derived parameters for the nominal and alternative models.
}
\label{tab2}
\centering
\begin{tabular}{llrrlr}
& & {\bf nominal} & alternative \\
comp. & var. & val. & val. & unit & $\sigma$ \\
\hline
\vrule height10pt width0pt
Ac1+Ac2 & $a_1          $ & $    54.032 $ & $    56.226 $ & $R_{\rm S}$ & $ 1    $ \\
Aa1+Aa2 & $a_2          $ & $   136.132 $ & $   128.052 $ & $R_{\rm S}$ & $ 10   $ \\
Ac+Aa   & $a_3          $ & $ 13053     $ & $ 12820     $ & $R_{\rm S}$ & $ 100  $ \\
Ac1+Ac2 & $e_1          $ & $     0.001 $ & $     0.0003$ & $1$         & $ 0.01 $ \\
Aa1+Aa2 & $e_2          $ & $     0.414 $ & $     0.353 $ & $1$         & $ 0.1  $ \\
Ac+Aa   & $e_3          $ & $     0.539 $ & $     0.304 $ & $1$         & $ 0.1  $ \\
Ac1     & $R_1          $ & $    16.626 $ & $    13.229 $ & $R_{\rm S}$ & $ 1    $ \\
Ac2     & $R_2          $ & $    11.975 $ & $    10.592 $ & $R_{\rm S}$ & $ 1    $ \\
Aa1     & $R_3          $ & $    28.286 $ & $    25.962 $ & $R_{\rm S}$ & $ 1    $ \\
Aa2     & $R_4          $ & $     4.340 $ & $     4.727 $ & $R_{\rm S}$ & $ 1    $ \\
Ac1     & $L_1          $ & $     0.220 $ & $     0.192 $ & $1$         & $ 0.05 $ \\
Ac2     & $L_2          $ & $     0.133 $ & $     0.189 $ & $1$         & $ 0.05 $ \\
Aa1     & $L_3          $ & $     0.635 $ & $     0.600 $ & $1$         & $ 0.02 $ \\
Aa2     & $L_4          $ & $     0.010 $ & $     0.017 $ & $1$         & $ 0.01 $ \\
\hline
\end{tabular}
\tablefoot{
$a$~denotes the semimajor axis,
$e$~the eccentricity,
$R$~the radius and
$L$~the relative luminosity in V band.
}
\end{table}

%%%%%%%%%%%%%%%%%%%%%%%%%%%%%%%%%%%%%%%%%%%%%%%%%%%%%%%%%%%%%%%%%%%%%%%%

\appendix

\section{Figures for the alternative model}\label{appendix}

\begin{figure}
\centering
\includegraphics[width=9cm]{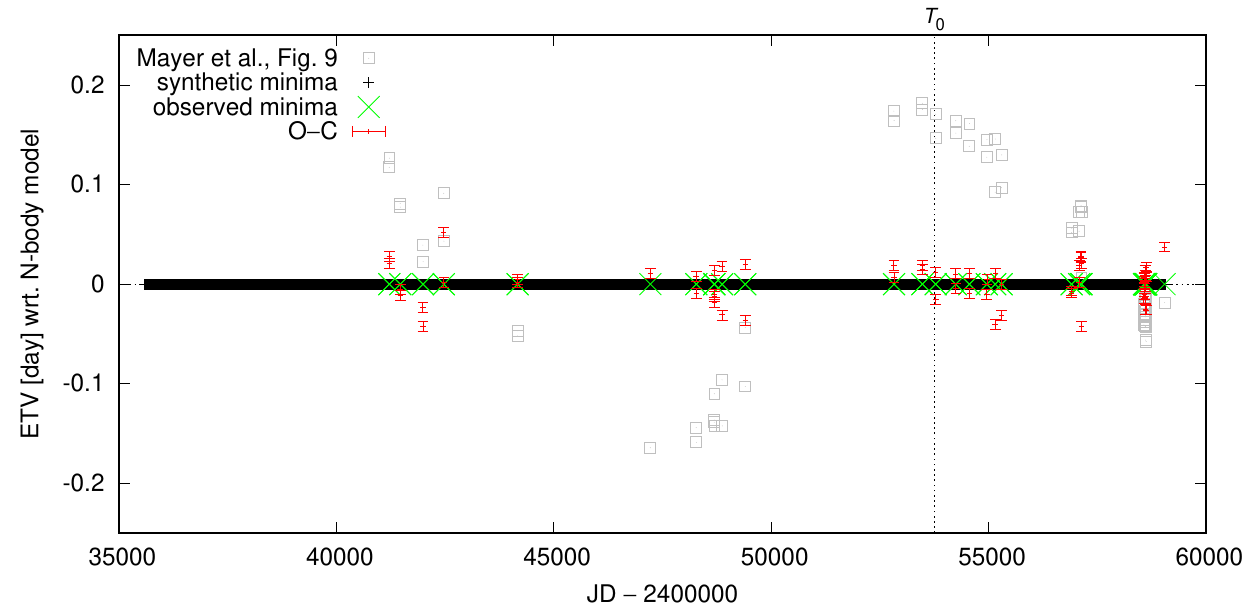}
\caption{Same as Fig.~\ref{chi2_TTV} for the alternative model.}
\label{qzcar_test32_alternative_chi2_TTV}
\end{figure}

\begin{figure}
\centering
\includegraphics[width=9cm]{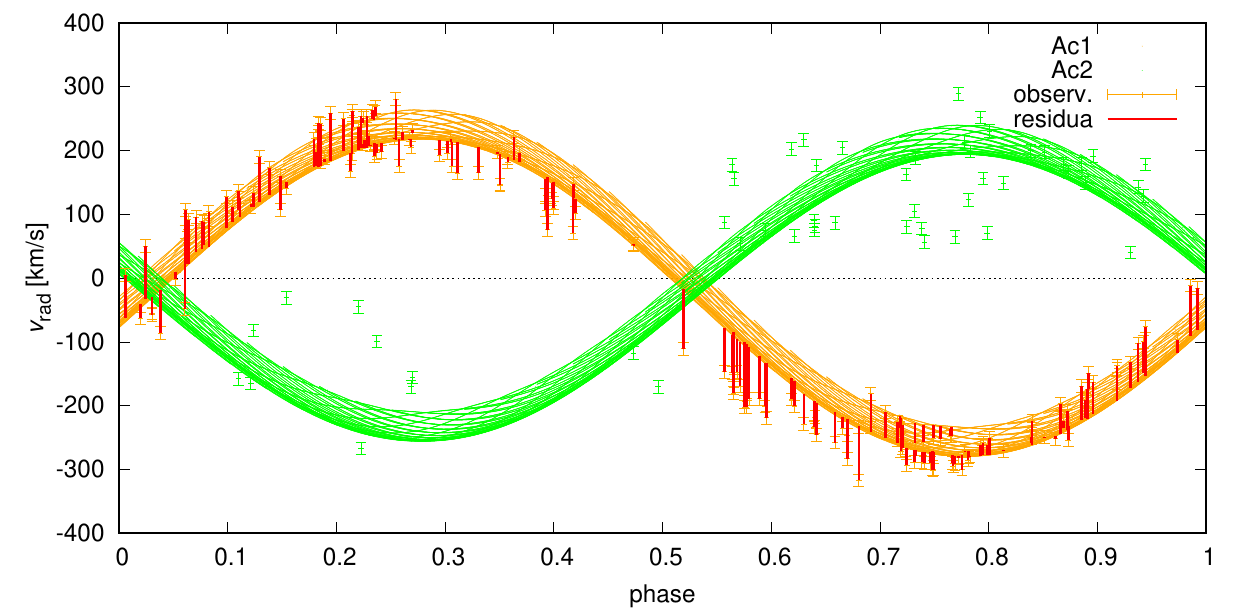}
\caption{Same as Fig.~\ref{chi2_RV1} for the alternative model.}
\label{qzcar_test32_alternative_chi2_RV1}
\end{figure}

\begin{figure}
\centering
\includegraphics[width=9cm]{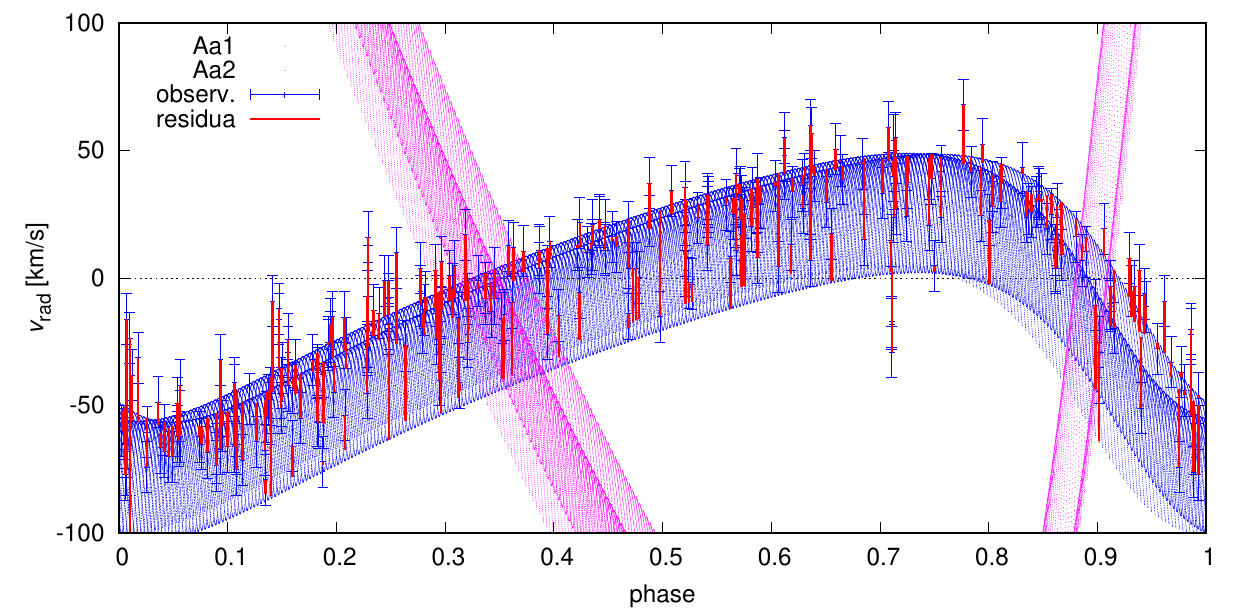}
\caption{Same as Fig.~\ref{chi2_RV3} for the alternative model.}
\label{qzcar_test32_alternative_chi2_RV3}
\end{figure}

\begin{figure}
\centering
\includegraphics[width=9cm]{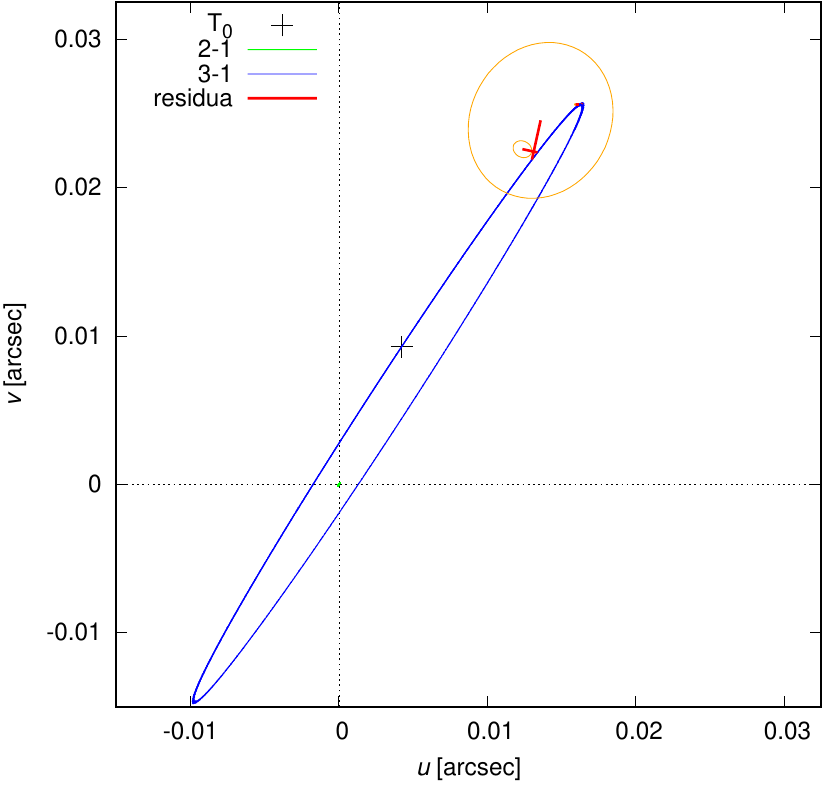}
\caption{Same as Fig.~\ref{chi2_SKY2} for the alternative model.}
\label{qzcar_test32_alternative_chi2_SKY2}
\end{figure}

\begin{figure*}
\centering
\includegraphics[width=18.0cm]{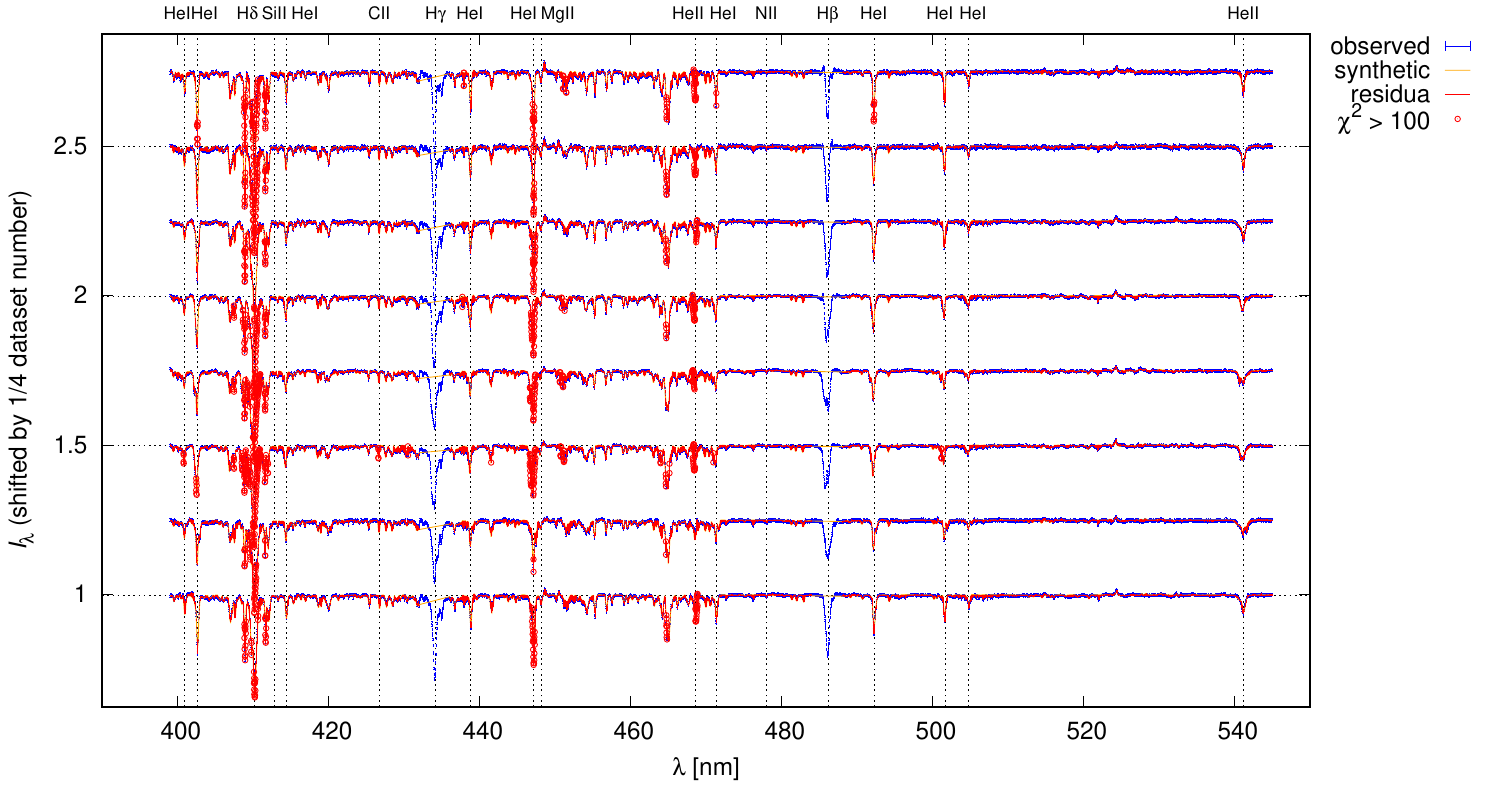}
\caption{Same as Fig.~\ref{chi2_SYN} for the alternative model.}
\label{qzcar_test32_alternative_chi2_SYN}
\end{figure*}

\begin{figure}
\centering
\includegraphics[width=9cm]{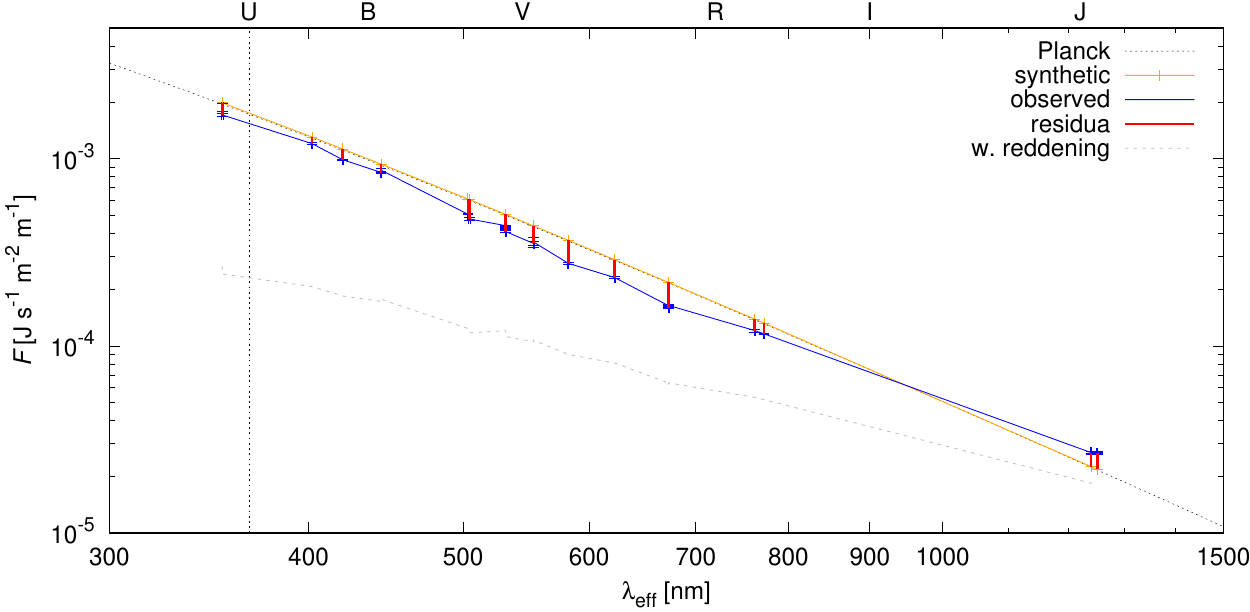}
\caption{Same as Fig.~\ref{chi2_SED} for the alternative model.}
\label{qzcar_test32_alternative_chi2_SED}
\end{figure}

\begin{figure}
\centering
\includegraphics[width=9.0cm]{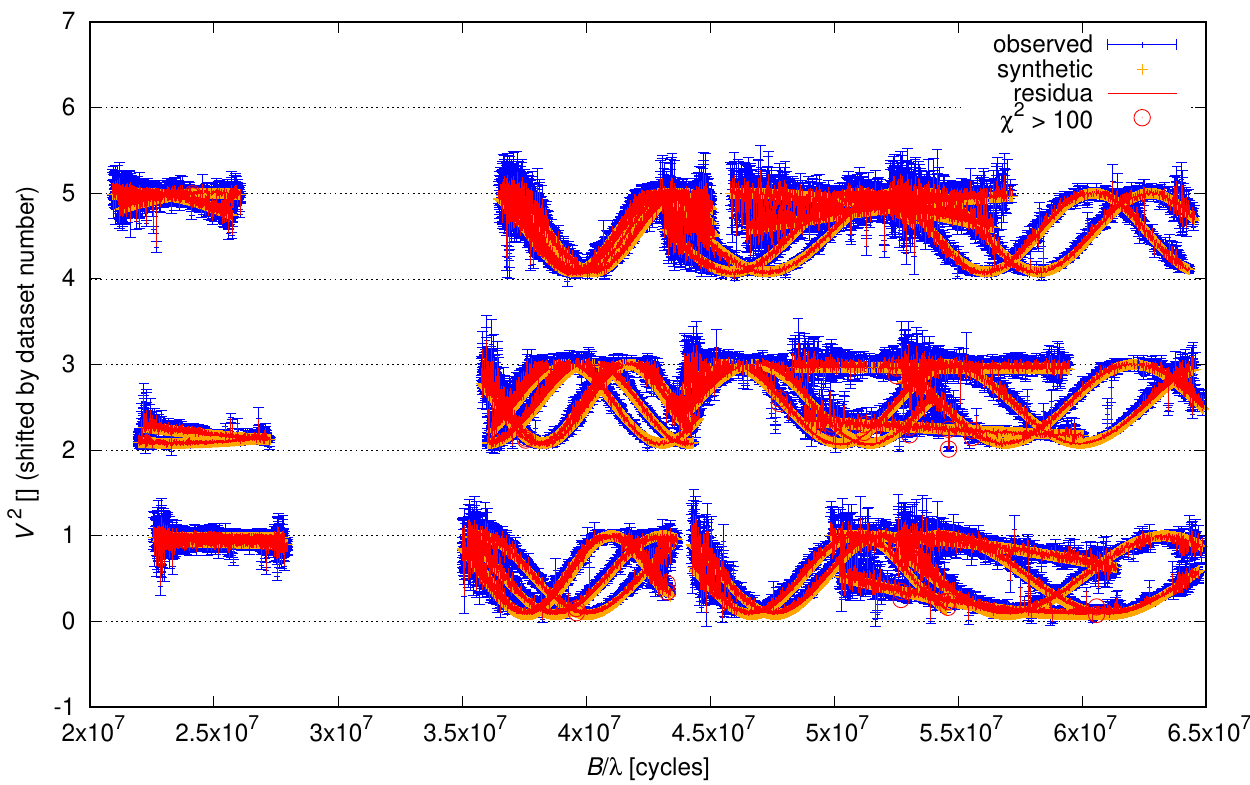}
\caption{Same as Fig.~\ref{chi2_VIS} for the alternative model.}
\label{qzcar_test32_alternative_chi2_VIS}
\end{figure}

\begin{figure}
\centering
\includegraphics[width=9cm]{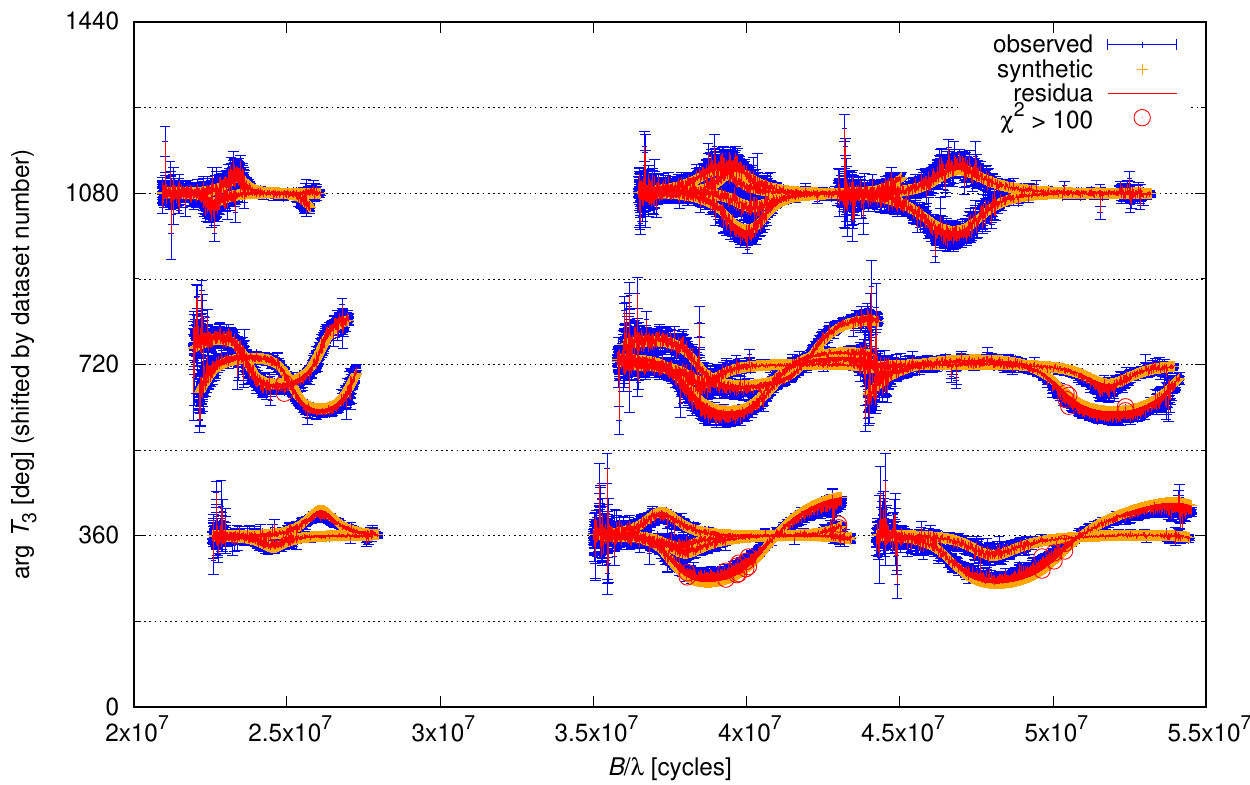}
\caption{Same as Fig.~\ref{chi2_CLO} for the alternative model.}
\label{qzcar_test32_alternative_chi2_CLO}
\end{figure}

For comparison, we show corresponding Figures for the alternative model
(Fig.~\ref{qzcar_test32_alternative_chi2_TTV} to Fig.~\ref{qzcar_test32_alternative_chi2_CLO}).
Generally, they are similar to the nominal model,
but they demonstrate,
how weights and additional constraints (for Ac2, Aa2 components)
can change the best-fit solution.

\end{document}